\documentclass[
 superscriptaddress,
 amsfonts,amsmath,amssymb,
 aps,prx,twocolumn,
 floatfix
]{revtex4-2}

\usepackage{graphicx}% Include figure files
\usepackage{dcolumn}% Align table columns on decimal point
\usepackage{bm}% bold math
\usepackage[dvipsnames]{xcolor}

\usepackage[caption=false]{subfig}
\usepackage{orcidlink}
\usepackage{cleveref}
\usepackage{alphabeta}
\usepackage{physics}
\usepackage{hyperref}

\DeclareMathOperator{\sinc}{sinc}

\DeclareMathOperator{\diag}{diag}
\let\var\relax
\DeclareMathOperator{\var}{Var}

\newcommand{\mnorm}[1]{\left\lVert#1\right\rVert}

\hypersetup{
    colorlinks=true,
    linkcolor=blue,
    citecolor=blue,      
    urlcolor=blue
    }

\begin{document}

\title{Universal cooling of quantum systems via randomized measurements }

\author{Josias Langbehn}
\thanks{These authors contributed equally to this work.}
\affiliation{Freie Universität Berlin, Fachbereich Physik and Dahlem Center for Complex Quantum Systems, Arnimallee 14, 14195 Berlin}

\author{George Mouloudakis\,\orcidlink{0000-0002-2591-5763}}
\thanks{These authors contributed equally to this work.}
\affiliation{Freie Universität Berlin, Fachbereich Physik and Dahlem Center for Complex Quantum Systems, Arnimallee 14, 14195 Berlin}

\author{Emma C. King\,\orcidlink{0000-0002-6696-3235}}
\affiliation{Theoretical Physics, Department of Physics, Saarland University, 66123 Saarbrücken, Germany}

\author{Rapha\"el Menu\,\orcidlink{0000-0001-7641-9922}}
\affiliation{Theoretical Physics, Department of Physics, Saarland University, 66123 Saarbrücken, Germany}

\author{Igor Gornyi\,\orcidlink{0000-0002-9253-6691}}
\affiliation{Institute for Quantum Materials and Technologies and Institut für Theorie der Kondensierten Materie, Karlsruhe Institute of Technology, Karlsruhe 76131, Germany}

\author{Giovanna Morigi\,\orcidlink{0000-0002-1946-3684}}
\affiliation{Theoretical Physics, Department of Physics, Saarland University, 66123 Saarbrücken, Germany}

\author{Yuval Gefen}
\affiliation{Department of Condensed Matter Physics, Weizmann Institute of Science, Rehovot 7610001, Israel}

\author{Christiane P. Koch\,\orcidlink{0000-0001-6285-5766}}
\email{christiane.koch@fu-berlin.de}
\affiliation{Freie Universität Berlin, Fachbereich Physik and Dahlem Center for Complex Quantum Systems, Arnimallee 14, 14195 Berlin}

\date{\today}

\begin{abstract}
Designing cooling protocols is believed to require knowledge of the system spectrum. In contrast, cooling in nature occurs whenever the system is coupled to a cold bath. How does nature know how to cool? A natural cold bath can be mimicked with a reservoir of ``meter'' qubits that are initialized in their ground state. We show that a quantum system can be cooled without knowledge of system details when system-meter interactions and meter splittings are chosen randomly. For sufficiently small interaction strengths and long interaction times, the protocol ensures that resonant energy-exchange processes, leading to cooling, dominate over heating. Effectively, the dynamics is then captured by the rotating-wave approximation, which we identify as the basic mechanism for robust and scalable cooling of complex quantum systems through generic, structure-independent protocols. This offers a versatile universal framework for controlling quantum matter far from equilibrium, in particular, for quantum computing and simulation.
\end{abstract}

\maketitle

\section{Introduction} \label{sec:introduction}

The development of efficient techniques to cool quantum systems has been instrumental for the emergence of quantum technologies. 
Laser cooling of atoms~\cite{cohen-tannoudji_1998_Rev.Mod.Phys._NobelLectureManipulating,phillips_1998_Rev.Mod.Phys._NobelLectureLaser,chu_1998_Rev.Mod.Phys._NobelLectureManipulation} paved the way for many identical quantum systems to be prepared in a pure quantum state. Extensions of laser cooling to the sideband and sympathetic cooling were pivotal to leveraging entanglement for a practical advantage in quantum logic spectroscopy \cite{wineland_2013_Rev.Mod.Phys._NobelLectureSuperposition}.
The necessity to freeze out all the degrees of freedom that are not relevant to the encoding and processing of quantum information is even more imperative in solid-state platforms. Superconducting qubits, for example, are kept in dilution refrigerators to ensure quantum coherence~\cite{girvin_2014_QuantumMachines:MeasurementandControlofEngineeredQuantumSystems_CircuitQEDSuperconducting}.

In addition to maintaining sufficiently long lifetimes of quantum states, the operation of quantum hardware also requires the ability to prepare a set of qubits in a non-trivial fiducial pure state~\cite{divincenzo_2000_Fortschr.Phys._PhysicalImplementationQuantum} and manipulate this state, avoiding unwanted excitations or decoherence. This represents a major factor limiting the speed and duration of coherent operations for most architectures. 
Clearly, present-day quantum platforms would benefit from the advancement of \textit{universal} cooling strategies, since noise and crosstalk due to uncontrolled interactions between qubits severely limit the reachable circuit depths, as well as the fidelity of the engineered states \cite{preskill_2018_Quantum_QuantumComputingNISQ, bharti_2022_Rev.Mod.Phys._NoisyIntermediatescaleQuantum}.  
Further, cooling could be useful for quantum algorithms that rely on measuring the energy of the final state, such as variational quantum algorithms \cite{cerezo_2021_NatRevPhys_VariationalQuantumAlgorithms,wecker_2015_Phys.Rev.A_ProgressPracticalQuantum,alexeev_2024_FutureGenerationComputerSystems_QuantumcentricSupercomputingMaterials}.

One difficulty in qubit reset and cooling in general is the export of entropy ~\cite{alicki_2018_ThermodynamicsintheQuantumRegime_IntroductionQuantumThermodynamics} --- it requires an environment into which entropy can be discarded. A possible approach to achieve this is by weakly coupling a ``natural'' bath, kept at low temperature, to the system. Other established techniques can render the entropy export more controllable and efficient. For example, the fastest way to reset a qubit is resonant energy exchange with an auxiliary qubit that is then discarded~\cite{ticozzi_2014_SciRep_QuantumResourcesPurification,basilewitsch_2021_Phys.Rev.Research_FundamentalBoundsQubit}, which is, however, based on knowledge about the system, here the relevant energy gap. Similarly, laser cooling relies on (near-) resonant excitation of optical transitions in a closed cycle~\cite{cohen-tannoudji_1998_Rev.Mod.Phys._NobelLectureManipulating,phillips_1998_Rev.Mod.Phys._NobelLectureLaser,chu_1998_Rev.Mod.Phys._NobelLectureManipulation}, which becomes increasingly difficult for more complex systems and has hampered, e.g., the laser cooling of molecules~\cite{fitch_2021_ChapterThreeLasercooled}. 
In general, the cooling efficiency of a quantum system relies on the design of scattering processes that lead to energy extraction and overcome all other unwanted incoherent processes. This becomes challenging with increasing system complexity, where uncontrollable couplings with detrimental environments may be present or ergodicity may be broken due to symmetries.
Take the example of cooling many-body quantum systems relevant to quantum simulations and quantum annealing. Not only are their spectra increasingly complex, but, by construction, generically unknown, making it difficult to devise suitable cooling schemes.
Dissipative state preparation protocols~\cite{diehl_2008_Nat.Phys_QuantumStatesPhases, diehl_2011_NaturePhys_TopologyDissipationAtomic, verstraete_2009_NaturePhys_QuantumComputationQuantumstate, zhan_2025_RapidQuantumGround, lin_2025_APLComput.Phys._DissipativePreparationManybody,morigi_2015_Phys.Rev.Lett._DissipativeQuantumControl,hagan_2025_ThermodynamicCostIgnorance} do require some knowledge of the system's spectrum, such as the spectral gap.
Likewise, measurement-based steering of quantum states which drives the system towards a predesignated target state by quasi-local measurements~\cite{roy_2020_Phys.Rev.Research_MeasurementinducedSteeringQuantum,  kumar_2020_Phys.Rev.Res._EngineeringTwoqubitMixed, herasymenko_2023_PRXQuantum_MeasurementDrivenNavigationManyBody, volya_2024_IEEETrans.QuantumEng._StatePreparationQuantum, morales_2024_Phys.Rev.Research_EngineeringUnsteerableQuantum} also requires knowledge of the system's degrees of freedom (and possibly Hamiltonian). 

\begin{figure}[tb]
    \centering
    \includegraphics[width=0.95\linewidth]{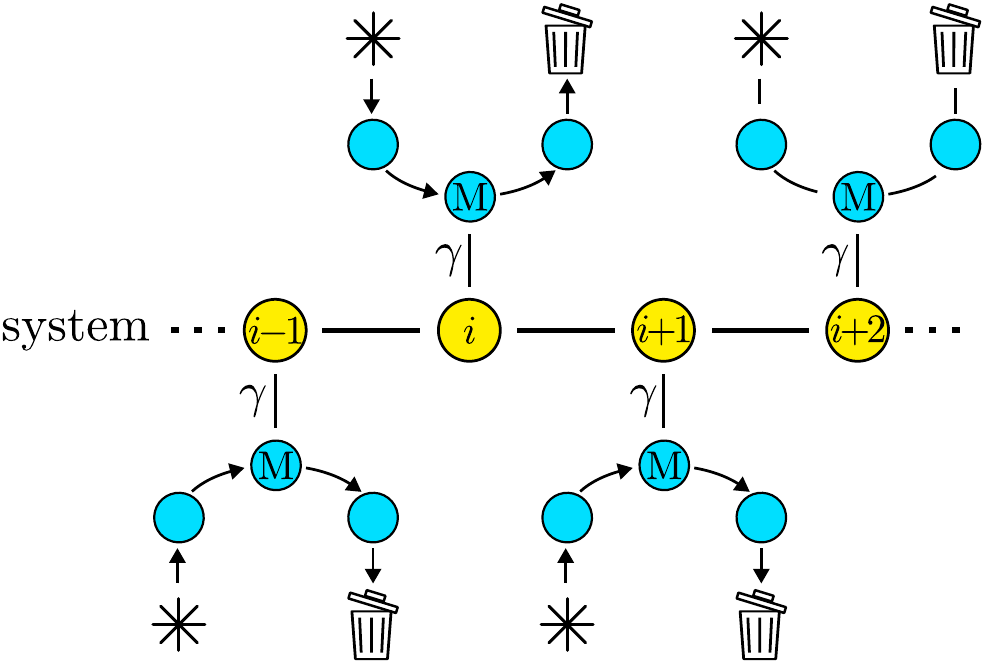}
    \caption{\textbf{Universal cooling via repeated, stochastically chosen interactions with meter qubits:} Each constituent of the system is coupled to a meter qubit with coupling strength $\gamma$. The meter qubits are initialized in their ground states and interact sequentially with the system for time $t_M$ after which they are discarded.}
    \label{fig:schematic}
\end{figure}

\nocite{englert_2002_CoherentEvolutioninNoisyEnvironments_FiveLecturesDissipativea}
These examples underline the challenge of devising cooling protocols with practically no knowledge of the system at hand. Here we address this problem by leveraging an analogy with nature: Why can a generic quantum system be cooled by a low-temperature bath, irrespective of its internal Hamiltonian?
The strategy for mimicking a cold bath to cool, or, more precisely, to thermalize a many-body quantum system is illustrated in Fig.~\ref{fig:schematic}: The cold bath, acting as a sink for both entropy and energy, is represented by a set of auxiliary ``meter'' qubits that are initialized in their ground state. They are coupled locally with system degrees of freedom (depicted in yellow in Fig.~\ref{fig:schematic}) and, after a fixed interaction (or ``measurement'') time $t_M$, are discarded and replaced with identically initialized meters. The process is repeated until the steady-state energy is attained with a given fidelity~\footnote{We refer to the auxiliary qubits as the \emph{measurement} or \emph{meter} qubits since discarding them after their interaction with the system is equivalent to performing a projective measurement without keeping track of the outcome~\cite{englert_2002_CoherentEvolutioninNoisyEnvironments_FiveLecturesDissipativea} (so-called ``blind measurement''~\cite{roy_2020_Phys.Rev.Research_MeasurementinducedSteeringQuantum})}. 
This scheme has been turned into practical protocols in the context of measurement-induced dynamics
\cite{
 zhao_2019_Phys.Rev.A_MeasurementDrivenAnalog, roy_2020_Phys.Rev.Research_MeasurementinducedSteeringQuantum,
kumar_2020_Phys.Rev.Res._EngineeringTwoqubitMixed, 
noel_2022_Nat.Phys._MeasurementinducedQuantumPhases,
menu_2022_Phys.Rev.Research_ReservoirengineeringShortcutsAdiabaticity, 
harrington_2022_Nat.RevPhys_EngineeredDissipationQuantum, 
koh_2023_Nat.Phys._MeasurementinducedEntanglementPhase, 
googlequantumaiandcollaborators_2023_Nature_MeasurementinducedEntanglementTeleportation, 
cubitt_2023_DissipativeGroundState,  
herasymenko_2023_PRXQuantum_MeasurementDrivenNavigationManyBody, 
smith_2023_PRXQuantum_DeterministicConstantDepthPreparation, 
sriram_2023_Phys.Rev.B_TopologyCriticalityDynamically, 
lavasani_2023_Phys.Rev.B_MonitoredQuantumDynamics,
matthies_2024_Quantum_ProgrammableAdiabaticDemagnetization, 
volya_2024_IEEETrans.QuantumEng._StatePreparationQuantum,
puente_2024_Quantum_QuantumStatePreparation,
chen_2024_Quantum_EfficientPreparationAKLT},
dissipative state preparation
~\cite{terhal_2000_Phys.Rev.A_ProblemEquilibrationComputation,
pielawa_2007_Phys.Rev.Lett._GenerationEinsteinPodolskyRosenEntangledRadiation,kraus_2008_Phys.Rev.A_PreparationEntangledStates,
diehl_2008_Nat.Phys_QuantumStatesPhases,
verstraete_2009_NaturePhys_QuantumComputationQuantumstate,
diehl_2011_NaturePhys_TopologyDissipationAtomic,
rojan_2014_Phys.Rev.A_ArbitraryquantumstatePreparationHarmonic,morigi_2015_Phys.Rev.Lett._DissipativeQuantumControl,
brown_2022_NatCommun_TradeOfffreeEntanglement, 
metcalf_2022_QuantumSci.Technol._QuantumMarkovChain,
kitzman_2023_NatCommun_PhononicBathEngineering,
langbehn_2024_PRXQuantum_DiluteMeasurementInducedCooling,
mi_2024_Science_StableQuantumcorrelatedManybody, ding_2024_Phys.Rev.Research_SingleancillaGroundState,
lloyd_2025_PRXQuantum_QuasiparticleCoolingAlgorithms,
kishony_2025_Phys.Rev.Lett._GaugedCoolingTopological,
zhan_2025_RapidQuantumGround,
lin_2025_APLComput.Phys._DissipativePreparationManybody,lloyd_2025_PRXQuantum_QuasiparticleCoolingAlgorithms},
heat-bath algorithm cooling
~\cite{boykin_2002_Proc.Natl.Acad.Sci.U.S.A._AlgorithmicCoolingScalable,fernandez_2004_Int.J.QuantumInform._ALGORITHMICCOOLINGSPINS,
schulman_2007_SIAMJ.Comput._PhysicalLimitsHeatBath,
linden_2010_Phys.Rev.Lett._HowSmallCan,
park_2016_ElectronSpinResonanceESRBasedQuantumComputing_HeatBathAlgorithmic,fogarty_2016_Phys.Rev.A_OptomechanicalManybodyCooling,
metcalf_2020_Phys.Rev.Research_EngineeredThermalizationCooling,
zaletel_2021_Phys.Rev.Lett._PreparationLowEntropy,polla_2021_Phys.Rev.A_QuantumDigitalCooling,feng_2022_Phys.Rev.A_QuantumComputingCoherent,
maurya_2024_PRXQuantum_OnDemandDrivenDissipation,
uppalapati_2024_AntithermalizationHeatingCooling,marti_2025_Quantum_EfficientQuantumCooling,molpeceres_2025_Phys.Rev.Research_QuantumAlgorithmsCooling},
collision models
~\cite{scarani_2002_Phys.Rev.Lett._ThermalizingQuantumMachines, 
ciccarello_2017_QuantumMeas.QuantumMetrol._CollisionModelsQuantum,
cattaneo_2021_Phys.Rev.Lett._CollisionModelsCan,
ciccarello_2022_PhysicsReports_QuantumCollisionModels}, 
or quantum-feedback cooling \cite{magrini_2021_Nature_RealtimeOptimalQuantum,kamba_2022_Opt.Express_OpticalColdDamping,tebbenjohanns_2021_Nature_QuantumControlNanoparticle,sugiura_2025_QuantumFeedbackCooling,doherty_1999_Phys.Rev.A_FeedbackControlQuantum}. 
Yet, the analogy with cooling in nature has not been fully explored; in particular, these protocols have not been designed to be system-agnostic.   
Attempts to reduce the required knowledge of the spectrum include sweeping the splitting of the meter qubits~\cite{polla_2021_Phys.Rev.A_QuantumDigitalCooling} and randomizing coupling operators~\cite{ding_2024_Phys.Rev.Research_SingleancillaGroundState} or interaction times~\cite{molpeceres_2025_Phys.Rev.Research_QuantumAlgorithmsCooling}. These advances are important in view of practical implementations, but they do not answer the question how and why a system can be cooled without full knowledge of its spectral properties and symmetries.

Here, we establish stochasticity together with weak coupling as the general principle underlying universal cooling.
We show that, in order to be as system-agnostic as possible, both the splittings of the meter qubits and the system-meter interactions have to be chosen randomly. Then, for small couplings $\gamma$ and long interaction times $t_M$, the steady-state energy is proportional to $\gamma$ and inversely proportional to $t_M$.
In this regime, stochasticity ensures that resonant energy exchange, which leads to cooling, is favored over non-resonant processes that may result in heating, akin to the requirements at the basis of the rotating wave approximation in quantum optics.
Remarkably, two-body interactions---the simplest possible choice for the system-meter interaction---are sufficient for cooling. 
Despite the local coupling, the Kraus operators that we derive in the limit of weak coupling and long interaction times are effectively non-local, an expected key requirement for the cooling of a generic many-body quantum system~\cite{kraus_2008_Phys.Rev.A_PreparationEntangledStates, fogarty_2016_Phys.Rev.A_OptomechanicalManybodyCooling, ding_2024_Phys.Rev.Research_SingleancillaGroundState}. 
Our protocol allows for cooling without the need to tailor the procedure to any of the system's details.
The price for agnosticity is the longer cooling time that is larger than those in approaches based on, e.g., feedback \cite{herasymenko_2023_PRXQuantum_MeasurementDrivenNavigationManyBody} or learning~\cite{marti_2025_Quantum_EfficientQuantumCooling}. Moreover, the cooling times are determined not just by stochastic convergence but also by the minimal spectral gap, which sets the scale for weak coupling. At the same time, our protocol, on top of being closest to mimicking thermalization with a natural bath, is astonishingly simple to implement. 

\section{Cooling Protocol} \label{sec:cooling_protocol}

The setup, combining the system of interest $S$ and the meter qubits, evolves under the Hamiltonian
\begin{equation}
    H_\text{tot} = H_\text{S} + H_\text{SM} + H_\text{M}\,,
    \label{Htot-eq1}
\end{equation}
where $H_\text{S}$ and $H_\text{M}$ act on the system and meter qubits, respectively, and their coupling is given by $H_\text{SM}$. 
If the spectrum of $H_\text{S}$ is known, the cooling process can be optimized by tailoring $H_\text{SM}$ and $H_\text{M}$ accordingly. Instead, our work is motivated by the observation that when a system is immersed in a cold bath, it thermalizes to the bath's temperature even though the bath is not specifically tailored to the system, i.e., the bath is system-agnostic. To obtain a cooling protocol without making assumptions about the system other than that it consists of quantum degrees of freedom with local couplings, we average over both the form of the system-meter interaction $H_\text{SM}$ and the level splitting of the meter qubits. 

The cooling protocol is illustrated in Fig.~\ref{fig:schematic} for the example of a one-dimensional many-body quantum system. It consists of repeating the following steps. First, the meter qubits are prepared in the ground state of their corresponding Hamiltonian $H_\text{M}=(\omega_\text{M}/2)\,\tau_z$, where $\tau_z$ is the Pauli-$z$ operator of the meter and the level splitting $\omega_\text{M}$ is chosen at random. In the second step, the meters and the system interact locally for an interaction time $t_M$ where the form of the interaction $H_\text{SM}$ is also chosen randomly. Finally, after the interaction time, the system and the meter are decoupled, the meters are discarded, and the cycle is repeated starting from step 1 with a different random value of $\omega_\text{M}$ and a different form of the system-meter coupling.

\subsection{Asymptotic limit of random measurements}\label{subsec:equivalence}
After many iterations of the protocol, the system will approach the steady state. We now show that it is related to the averaged completely positive trace-preserving (CPTP) map $\mathcal{\bar{V}}$ for one iteration with randomized splittings and couplings: 
Initially, the system and all meter qubits are uncorrelated,
    $\rho(0)= \rho_\text{S}(0) \otimes |\vec{0}\rangle \langle \vec{0}|_\text{M}$
where $|\vec{0}\rangle_\text{M}$ is the product state of the meter ground states. 
We describe the system-meter interactions by the unitary $U(t) = \sum_{\vec{i},\vec{j}} U_{\vec{i},\vec{j}}(t) |\vec{i}\rangle \langle\vec{j}|_\text{M}$ where $|\vec{i}\rangle_\text{M}$ runs over the Hilbert space of the meters with $\vec{i}=(i_1,\ldots,i_N)$ and $i_k$ enumerates the basis states of the $k$-th meter out of the $N$ meters. The global state at time $t$ is $\rho(t)=U(t)\rho_\text{S}(0)\otimes |\vec{0}\rangle \langle \vec{0}|U^\dagger(t)$.
Tracing out the meters, the reduced state of the system after one iteration is given by
\begin{align}
    \rho_\text{S}(t_M) &= \sum_{\vec{i}} \bra{\vec{i}}_\text{M}U(t_M)\rho_\text{S}(0)\otimes\ketbra{\vec{0}}_\text{M} U^\dagger(t_M)\ket{\vec{i}}_\text{M} \nonumber\\
    &= \sum_{\vec{i}} U_{\vec{i}, \vec{0}}(t_M) \rho_\text{S}(0)U_{\vec{i},\vec{0}}^\dagger(t_M) 
    =\mathcal{V}\rho_\text{S}(0)\,,
    \label{eq:app_cptp_from_u}
\end{align}
where $\mathcal{V}$ is a CPTP map describing the incoherent dynamics emerging from the interaction with the meters.
To describe a stochastic average over many interactions, we parametrize the interaction $H_\text{SM}$ and choice of meter splitting of the $n$-th iteration in terms of a vector $\vec{r}_n$, such that the CPTP map, mapping step $n-1$ to step $n$, becomes $\mathcal{V}\left(\vec{r}_n\right)$.
Averaging over all possible choices of system-meter interaction and meter splittings for a single iteration gives the map
\begin{align}
    \bar{\mathcal{V}} &= \frac{1}{\mathcal{N}}\int d\vec{r}\ \mathcal{V}\left(\vec{r}\right)
\end{align}
with fixed variance and suitable normalization $\mathcal{N}$.
The averaged state of the system after $n$ iterations is then obtained by integrating the individual realizations
\begin{align}
    \rho_S\left(\vec{r}_n,\ldots,\vec{r}_1 \right)&=\mathcal{V}(\vec{r}_n)\ldots \mathcal{V}(\vec{r}_1) \rho_S(0) \label{eq:single_trajectory}
\end{align}
over all parameters $\vec{r}_n$: 
\begin{align}
    \bar{\rho}_\text{S}(n\,t_M)&\equiv \frac{1}{\mathcal{N}^n}\int d\vec{r}_n\ldots d\vec{r}_1 \mathcal{V}(\vec{r}_n) \ldots \mathcal{V}(\vec{r}_1) \rho_\text{S}(0) \nonumber\\
    &=\bar{\mathcal{V}}^n \rho_\text{S}(0)\,.\label{eq:app_cptp_average}
\end{align}
Therefore the average over many trajectories for an initial state $\rho_\text{S}(0)$ is equivalent to propagating $\rho_\text{S}(0)$ with the averaged CPTP map $\bar{\mathcal{V}}$.
Note that $\bar{\rho}_\text{S} (n\,t_M)$ is the mean value of the distribution of state trajectories after $n$ repetitions. We expect that individual trajectories converge to the average $\rho_S(\vec{r}_n,\ldots,\vec{r}_1) \rightarrow \bar{\rho}_S(n\,t_M)$ as $n\rightarrow\infty$, i.e., the variance of the distribution of states vanishes only as $n\rightarrow \infty$. Nevertheless, all steady-state properties can be extracted from $\bar{\mathcal{V}}$ by calculating its fixed point.
In other words, the system state at the asymptotic time $t = n t_M$ can be deduced, as $n\rightarrow\infty$, from the CPTP map for a \textit{single} iteration, averaged over all system-meter interactions. A study of individual trajectories and their dependence on the distribution of $\vec{r}_n$ is presented in Sec:~\ref{subsec:implementation}.

To first order in system-meter coupling strength, our protocol is equivalent to coupling the system to a bath composed of all possible meter realizations~\cite{cattaneo_2021_Phys.Rev.Lett._CollisionModelsCan,ciccarello_2022_PhysicsReports_QuantumCollisionModels}, as we also show in Appendix~\ref{app:equivalence_to_continuous_bath}.
Next, we will use the example of a two-level system to identify the conditions on the system-meter coupling strength $\gamma$ and interaction time $t_M$ that ensure cooling towards the ground state of $H_S$.

\subsection{Cooling a qubit} \label{subsec:results_qubit}

\begin{figure}
    \centering
    \includegraphics[width=1\linewidth]{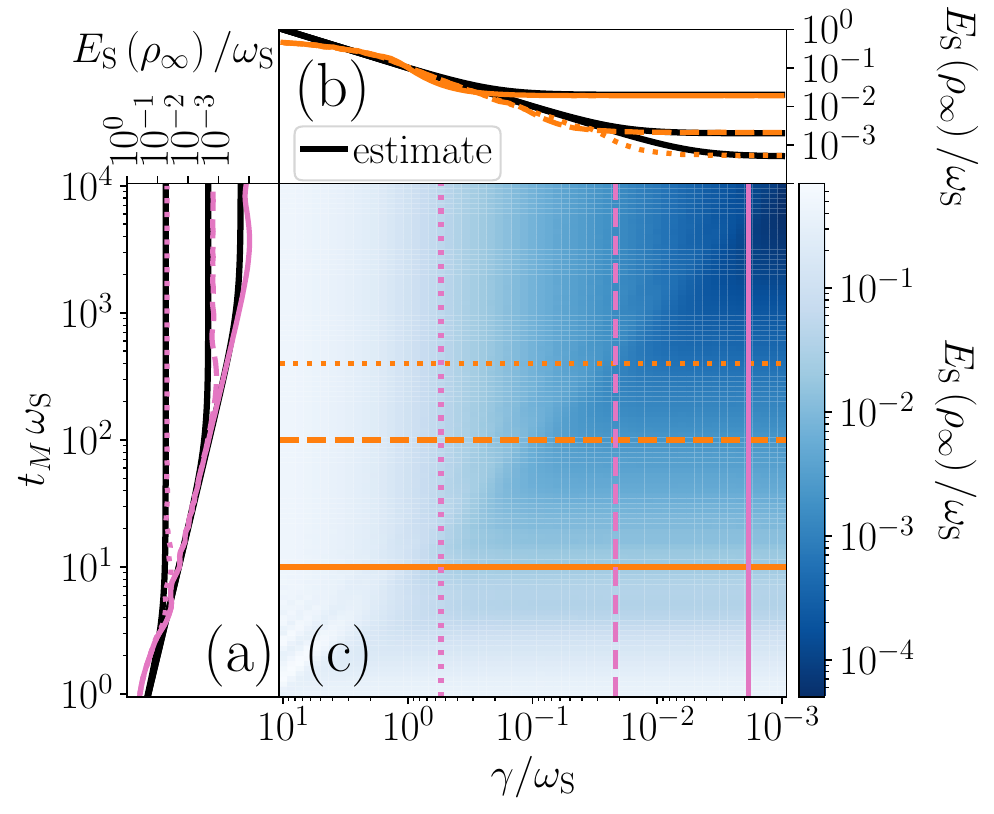}
    \caption{\textbf{Cooling a two-level system:} Steady-state energy $E_\text{S}(\rho_\infty)$ as a function of system-meter coupling strength $\gamma$ and interaction time $t_M$ with cuts along $t_M$ and $\gamma$ presented, respectively, in (a) and (b) by the corresponding (dotted, dashed, and solid) magenta and black curves. 
    The steady-state energy follows the estimate \eqref{eq:es_estimate} for $\omega_\text{S}\gg 1/t_M$ and $\omega_\text{S}\gg \gamma$, as indicated by the black curves.}
    \label{fig:tls_3panels}
    \refstepcounter{subfigure}
    \label{fig:tls_3panels_a}
    \refstepcounter{subfigure}
    \label{fig:tls_3panels_b}
    \refstepcounter{subfigure}
    \label{fig:tls_3panels_c}
\end{figure}

To consider the cooling of a single qubit, we write its Hamiltonian as 
\begin{align} H_\text{S} &= \frac{\omega_\text{S}}{2} \, \vec{v} \cdot \vec{\sigma}\,, 
\end{align}
where $\omega_\text{S}$ is the system qubit's level splitting, $\vec{v}$ the quantization axis in the Bloch sphere, and $\vec{\sigma}$ the vector of Pauli matrices. We choose
\begin{align}
    H_\text{SM} &= \gamma\, \sigma (\theta,\phi) \otimes \tau_x, 
\end{align}
where $\sigma(\theta,\phi)$ is a Pauli matrix representing quantization along the direction corresponding to angles $\theta$ and $\phi$ on the Bloch sphere.
The Hamiltonian of the meter qubits is 
\begin{equation}
 H_\text{M} = \frac{\omega_\text{M}}{2}\, \tau_z, 
\end{equation} 
where $\omega_\text{M}$ is the level splitting of the meter (given in the laboratory frame). We choose, without loss of generality, $\omega_\text{M}>0$, such that the meter's ground state is $|\!\!\downarrow\rangle$. 
The interaction between the system qubit and the meter leads to a CPTP map $\mathcal{V}(\theta,\phi,\omega_\text{M})$ that describes the evolution of the system from iteration $n$ to $n+1$. The steady state is obtained by averaging $\mathcal{V}(\theta, \phi, \omega_\text{M})$ over the Bloch sphere and uniformly over $\omega_\text{M} \in [0.1 \abs{\omega_\text{S}},3 \abs{\omega_\text{S}}]$. This choice of $\omega_\text{M}$ %only requires 
assumes an estimate of the spectral radius of $H_\text{S}$, such as to cover all possible transitions. Without this knowledge, $\omega_\text{M}$ can instead be sampled from a tailed distribution covering all frequencies. While it may come at the expense of a larger number of iterations, it guarantees cooling. Importantly, it is the initialization of
the meters in their ground state which ensures that the meters can absorb energy and mirror a bath at zero temperature.

Figure~\ref{fig:tls_3panels} shows the system energy $E_\text{S}$ in steady state $\rho_\infty$ as a function of the coupling strength $\gamma$ and the interaction time $t_M$. It should be compared to the ground-state energy, with the energy scale chosen such that the ground state energy is zero.
A significant reduction of the qubit's energy compared to the initial value is observed, which indicates that the system is efficiently cooled down by randomized interactions with the meter.
For large $t_M$ and small $\gamma$, the system energy deviates from the ground-state energy by approximately $10^{-3}\omega_\text{S}$, which we refer to as the error. 
Notably, in Fig.~\ref{fig:tls_3panels_a}, the steady-state energy decreases as $1/t_M$ as long as $\omega_\text{S}\gg 1/t_M \gg \gamma$. When $1/t_M$ becomes comparable to $\gamma$, the energy as a function of $1/t_M$ reaches a saturation limit, cf.\ the different onsets of the saturation regimes for the three different values of $\gamma$.
Similarly, Fig.~\ref{fig:tls_3panels_b} shows a decrease of the steady-state energy as a function of $\gamma$ as long as $\gamma \gg 1/t_M$. Once $\gamma$ becomes comparable to $1/t_M$, the energy as a function of $\gamma$ saturates, as illustrated by the onset of the saturation regime for three values of $1/t_M$.

The dependence of the steady-state energy on $\gamma$ and $1/t_M$ observed in Figs.~\ref{fig:tls_3panels_a} and \ref{fig:tls_3panels_b} implies that cooling the system towards its ground state is possible with an arbitrarily low error by increasing $t_M$ and reducing $\gamma$ at the same time. Indeed, generically, for $\omega_\text{S} t_M\rightarrow\infty$ and $\gamma/\omega_\text{S} \rightarrow0$, the system's steady state tends to the ground state (see Appendix~\ref{app:ground_state_steady_state}).
For the regime of weak couplings ($ \gamma \ll \omega_\text{S}$) and long interaction times ($1/t_M \ll \omega_\text{S}$), we estimate the steady-state energy as
\begin{align}
    E_\text{S} \approx E_{\text{est}} \equiv \sqrt{(\gamma/2)^2 + 1/t_M^2}, \label{eq:es_estimate}
\end{align}
and find qualitative agreement with numerical simulation, as indicated by the black curves in Figs.~\ref{fig:tls_3panels_a} and \ref{fig:tls_3panels_b}.
Cooling of a qubit for different interaction models as well as randomized interactions has been studied in \cite{prositto_2025_QuantumSci.Technol._EquilibriumNonequilibriumSteady} suggesting similar conclusions. However there the authors require resonance of system and meter such that the proposal is not fully system agnostic.

\subsection{Why cooling works} \label{subsec:RWA_qubit}

We now show that weak coupling and long interaction times ensure an effective implementation of resonant energy exchange between the system and meters and, thus, universal cooling. It is akin to effectively implementing a rotating-wave approximation (RWA). To see this, assume for a moment that the system's quantization axis is along $z$, 
\begin{equation} 
    H_\text{S} = \frac{\omega_\text{S}}{2}\, \sigma_z\,,
    \label{eq:HSsigmaz}
\end{equation}
and the system-meter coupling is given by
\begin{equation}
    H_\text{SM} = \gamma\, \sigma_x \otimes \tau_x\,.
    \label{eq:HSMxx}
\end{equation}
This coupling comprises both co-rotating and counter-rotating terms,
\begin{equation}
    H_\text{SM} =  \underbrace{\frac{\gamma}{2}\big(\sigma_+ \tau_- + \sigma_-\tau_+\big)}_{H_{+-}} + \underbrace{\frac{\gamma}{2}\big(\sigma_+\tau_+ + \sigma_- \tau_-\big)}_{H_{++}}. \label{eq:h_xx_interaction}
\end{equation}
where $\sigma_+ = \ketbra{\uparrow}{\downarrow},\,\sigma_-=\sigma_+^\dagger$, and equivalently for the meter operators $\tau_\pm$.
For $\omega_\text{S} > 0$ in Eq.~\eqref{eq:HSsigmaz}, the counter-rotating term is $H_{++}$ and the co-rotating term is $H_{+-}$, but their roles are swapped for the opposite sign of $\omega_S$.
Under the RWA, one neglects the counter-rotating term and retains only the co-rotating one.

The co-rotating interaction necessarily increases the meter’s energy by transferring energy from the system (or keeps it constant in the rare case of the interaction time matching a full return of the excitation) since the meter is initialized in its ground state. Repeated meter resets then lead to cooling of the system. Indeed, it is well-known that a (partial) swap generated by the co-rotating terms is the unique unitary that leads to thermalization irrespective of initial system and meter states~\cite{scarani_2002_Phys.Rev.Lett._ThermalizingQuantumMachines,ziman_2002_Phys.Rev.A_DilutingQuantumInformation}.
In contrast, the counter-rotating interaction describes the simultaneous excitation (or de-excitation) of both the system and the meter. This can lead to an increase in the energy of both parties due to the accumulation of negative energy in the interaction term $H_\text{SM}$. 

Incidentally, the single-qubit steering protocol of Ref.~\cite{roy_2020_Phys.Rev.Research_MeasurementinducedSteeringQuantum}  realizes the counter-rotating interaction with $H_\text{SM}= H_{+-}$, assuming $\omega_\text{S}<0$ (while the co-rotating term is absent). Then the target state of the system qubit is identical to the initial state of the meter. While $|\!\!\downarrow\rangle$ is the ground state for the meter, it is the excited state for the system with $\omega_\text{S}<0$. As a result, the steering protocol leads to heating rather than cooling of the system qubit; see details in Appendix~\ref{App:steering}, as well as Ref.~\cite{uppalapati_2024_AntithermalizationHeatingCooling}. 

This example is important as it contradicts the naive expectation that fresh meters always form a zero-temperature bath that cools the system. It also elucidates the crucial role of resonant energy exchange for cooling when both co- and counter-rotating terms are present in the system-meter coupling. 
Indeed, within the RWA, the detrimental process of the ``counter-rotating heating'' is suppressed to first order in $\gamma$. We thus conjecture that our protocol is based on effectively realizing system-meter interactions in the RWA. The RWA is valid under the following conditions (see also Appendix~\ref{app:rwa_validity}):
\begin{subequations}\label{eq:RWA}
\begin{alignat}{2}
    \frac{|\abs{\omega_\text{S}}-\omega_\text{M}|}{\abs{\omega_\text{S}} + \omega_\text{M} } &\ll 1 \quad & \text{(near resonance),} \label{eq:rwa2_resonance}\\
        \gamma \ll \abs{\omega_\text{S}} &+ \omega_\text{M} \quad & \text{(weak coupling),} \label{eq:rwa1_weak_coupling}\\
    1/\left(\abs{\omega_\text{S}} + \omega_\text{M}\right) &\ll t_M \qquad & \text{(long interaction time).} \label{eq:rwa3_time}
\end{alignat}
\end{subequations}
Note that the resonance condition \eqref{eq:rwa2_resonance} together with Eq.~\eqref{eq:rwa1_weak_coupling} implies the more familiar weak coupling condition $\gamma \ll \abs{\omega_\text{S}}$ (up to an irrelevant factor of $2$). Equation \eqref{eq:rwa1_weak_coupling} highlights the fact that the RWA is a perturbative treatment of the coupling between system and meter, i.e., the coupling strength $\gamma$ has to be compared to the uncoupled Hamiltonian $H_\text{S}+H_\text{M}$, cf. Appendix~\ref{app:rwa_validity}.
With the energy gap $\omega_\text{S}$ unknown and $\omega_\text{M}$ chosen randomly, the resonance condition~\eqref{eq:rwa2_resonance} will only be fulfilled sometimes. This suggests that the requirements of weak coupling, Eq.~\eqref{eq:rwa1_weak_coupling}, and long interaction times, Eq.~\eqref{eq:rwa3_time}, already identified in our discussion of Fig.~\ref{fig:tls_3panels} above, are sufficient to effectively ensure the RWA. We analyze this in detail below.

\subsection{Role of co-rotating and counter-rotating terms in cooling a qubit} \label{subsec:single_qubit}

\begin{figure*}
    \centering
    \includegraphics[width=1\linewidth]{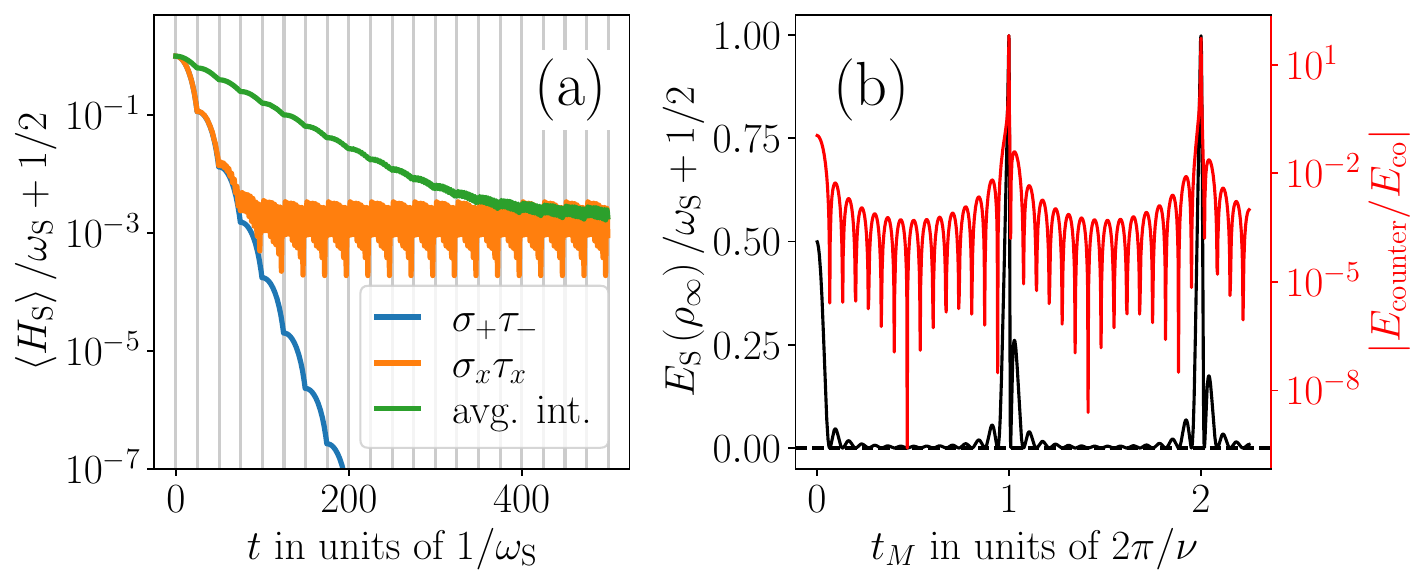}
    \vspace{-1\baselineskip}
    \caption{\textbf{Dynamics of cooling a two-level system via repeated interactions:}
    (a) System energy as a function of time for three different types of system-meter interaction corresponding to different levels of knowledge about the system. Stochastically averaged interactions (green) are used in the case of no knowledge, whereas $\sigma_x\tau_x$ (black) and $\sigma_+\tau_-$ (blue) couplings imply, respectively, knowledge of the quantization axis and knowledge of both the quantization axis and the gap sign. Note that in general the system attains a \emph{stroboscopic} steady state, i.e, at $t=n t_M$ when the meter is reset (indicated by faint vertical lines). In between measurements, the system energy can still show oscillations.
    (b) Steady-state energy (black) and the ratio of contributions to the steady-state energy associated with the counter- and co-rotating interactions (red), displaying revivals of the co-rotating terms at integer multiples of $\nu t_M/2\pi$, where $\nu$ is an eigenfrequency of $H_\text{tot}$, Eq.~\eqref{eq:tls_nu}.
    Except for the close vicinity of these values, the RWA is valid since the co-rotating contributions to the system energy dominate over the counter-rotating ones. Hence, except for singular unfortunate choices of $t_M$, the system energy can be reduced, implying efficient cooling. 
    Parameters for both panels are $\omega_\text{S}>0$,  $\gamma=0.1\omega_\text{S}$, and $\omega_\text{M}=\omega_\text{S}$.}
    \label{fig:main}
    \refstepcounter{subfigure}
    \label{fig:knowledge_hierarchy}
    \refstepcounter{subfigure}
    \label{fig:rwa_ratio}
\end{figure*}

Figure~\ref{fig:knowledge_hierarchy} presents the time evolution of the system energy under repeated meter resets for different choices of $H_\text{SM}$, corresponding to different levels of knowledge about the system. 
Assuming full knowledge of the system (quantization axis as well as the sign of $\omega_\text{S}$, taken here $\omega_S>0$), one can engineer the system-meter interaction such that it coincides with the co-rotating term $ H_{+-}$ (represented by the blue curve in Fig.~\ref{fig:knowledge_hierarchy}). Not surprisingly, the result is very efficient cooling toward the ground state. 
Next, assume that the system's quantization axis is known, while the sign of $\omega_\text{S}$ is not. A detailed analytical derivation for this case is presented in Appendix~\ref{App:steering}. As discussed above, for $\omega_\text{S} < 0$, the interaction $H_{+-}$ would lead to heating of the system because the roles of co- and counter-rotating interactions are reversed. Therefore, to be able to cool the system for both $\omega_\text{S} > 0$ and $\omega_\text{S} < 0$, both $H_{+-}$ and $H_{++}$ must be present; in other words, the interaction must be of the form $\sigma_x \tau_x$ (black curve in Fig.~\ref{fig:knowledge_hierarchy}). The important feature here is that $\sigma_x$ is perpendicular to $\sigma_z$, i.e., $\sigma_x$ acting on one of the $\sigma_z$-eigenstates maps it to the other one. In contrast to perfect cooling (blue curve), the black curve does not approach the ground-state energy, but the residual system energy is much lower than the initial one, and this error can be made smaller by choice of $\gamma$ and $t_M$, according to Fig.~\ref{fig:tls_3panels} and in line with Eqs.~(\ref{eq:rwa1_weak_coupling},~\ref{eq:rwa3_time}).
Finally, if the quantization axis is also unknown, the best one can do is to choose the direction of the interaction randomly, cf. green curve in Fig.~\ref{fig:knowledge_hierarchy}. 
It is obtained by propagating the initial state $\rho_\text{S}(0)$ with the averaged CPTP map $\bar{\mathcal{V}}$ such that $\rho_\text{S}(n\,t_\text{M}+ t) = \bar{\mathcal{V}}(t)\bar{\mathcal{V}}^n(t_\text{M})\rho_\text{S}(0)$ for $0\leq t \leq t_\text{M}$. This is equivalent to an average over many trajectories for individual realizations $\vec{r}_n,\ldots,\vec{r}_1$ of the system meter interaction and meter splitting, cf. Eq.~\eqref{eq:app_cptp_average}. For randomly chosen interaction directions,
the co-rotating part of the interaction (with respect to the unknown system's quantization axis) is dominant at some steps of the protocol, which, on average, overcomes the unwanted heating at other steps. While the cooling rate for the green curve is lower than that for the protocols tailored to the system, the final energy eventually approaches the value attained for a known quantization axis (black curve in Fig.~\ref{fig:knowledge_hierarchy}).

\begin{figure*}[t]
    \centering
    \subfloat[]{\includegraphics[width=0.48\textwidth]{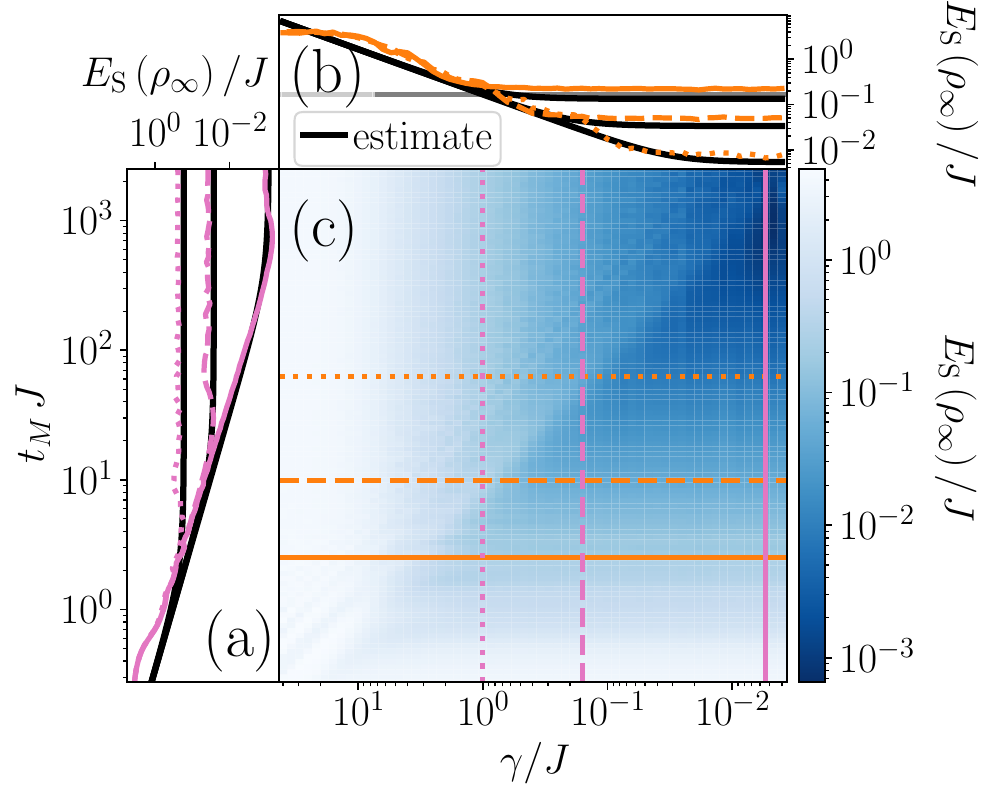}}
    \hfill
    \subfloat[]{\includegraphics[width=0.48\textwidth]{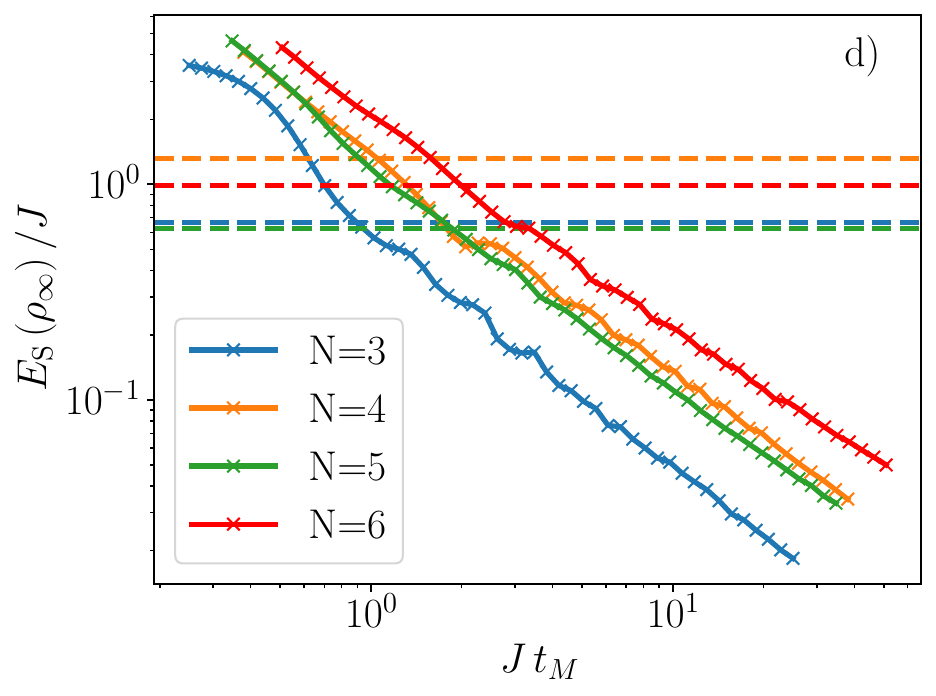}}
    \vspace{-1\baselineskip}
    \caption{\textbf{Cooling a Heisenberg chain:} (a-c) Steady-state energy as a function of system-meter interaction strength $\gamma$ and interaction time $t_M$ for $N=3$. As in the case of a two-level system (cf.\ Fig.~\ref{fig:tls_3panels}), cooling works best for weak interactions and long interaction times, where it follows the estimate \eqref{eq:es_estimate}.
    (d) Steady-state energy of the system as a function of interaction time $t_M$ for chain sizes $N=3,4,5,6$, all showing a scaling of the steady-state energy with $1/t_M$. The dashed lines indicate the lower bound on the energy obtainable by steering. Parameters are $\omega_\text{M}\in[0,1.1\mnorm{H_\text{S}}]$ and $\gamma=4 \times10^{-3} J$ in (d). The averaging of the CPTP map was done using Monte Carlo integration. The error due to the finite number of samples is of the order of $10^{-6}$ and thus not visible in this plot. All panels are for translation-invariant chains with open boundary conditions.}
    \label{fig:heisenberg}
    \refstepcounter{subfigure}
    \label{fig:heisenberg_3panels_a}
    \refstepcounter{subfigure}
    \label{fig:heisenberg_3panels_b}
    \refstepcounter{subfigure}
    \label{fig:heisenberg_3panels_c}
    \refstepcounter{subfigure}
    \label{fig:heisenberg_scaling}
\end{figure*}

Again, the residual system energy depends on the system-meter coupling strength and interaction time, cf. Fig.~\ref{fig:tls_3panels}, and can be made arbitrarily small.
This demonstrates that randomization of the system-meter coupling can efficiently cool a single qubit without knowing the sign of the gap and the quantization axis.
Note that in the case of $\sigma_x \tau_x$ or averaged coupling (black and green curves), the system energy in Fig.~\ref{fig:knowledge_hierarchy} shows oscillatory behavior between measurements. That is to say, the system converges \emph{stroboscopically} to the steady state at times $t=nt_\text{M}$ when the meter is reset. This poses no practical limitation because eventually we are interested in the system at long times, after it is decoupled from the meter.

Taking the interaction to be of the form $\sigma_x \tau_x$, the relation between cooling and the role of the co- and counter-rotating terms for $\omega_\text{S}>0$ is illustrated in Fig.~\ref{fig:rwa_ratio}. It compares the energy of the system's steady state as a function of interaction time $t_M$ (black curve, left y-axis labels) to the ratio $E_\text{counter}/E_\text{co}$ of the system energies associated with the counter-rotating and co-rotating interactions (red, right y-axis labels). The smaller the ratio, i.e., the more the ``co-rotating interaction energy'' dominates over its counter-rotating counterpart, the lower is the steady-state energy. In other words, the better the RWA is obeyed, the colder the system becomes.

Figure~\ref{fig:rwa_ratio} also shows that the steady-state energy depends on the choice of the interaction time $t_M$. In particular, for integer multiples of $\nu t_M/2\pi$, where $\nu$ is one of the eigenfrequencies of $H_\text{tot}$ [Eq.~\eqref{Htot-eq1};  see Appendix~\ref{App:steering}],
\begin{align}
    \nu &= \sqrt{\gamma^2 + \left(\omega_\text{S} - \omega_\text{M}\right)^2}\,, \label{eq:tls_nu}
\end{align}
``revivals'' of the counter-rotating contributions to the energy are observed. At these interaction times, the energy associated with the co-rotating processes is swapped from the system to the meter and back, such that $E_\text{co}$ becomes small in comparison to the counter-rotating contribution $E_\text{counter}$. This may lead to heating of the system. In practice, this does not pose an obstacle, since the effect occurs for relatively isolated values of $t_M$ only. Moreover, the position of the revivals depends on the specific interaction and meter splitting. Since we investigate stochastic choices of those parameters, on average, these peaks will be suppressed (the same would also happen for randomized values of $t_M$~\cite{molpeceres_2025_Phys.Rev.Research_QuantumAlgorithmsCooling}).
Looking ahead to many-body systems, one may wonder whether the density of the revival peaks in Fig.~\ref{fig:rwa_ratio} will increase since more system transition frequencies may lead to the revival condition to be fulfilled more often. However, actual revivals will only occur for meter splittings that are resonant to one of those system transition frequencies. The situation is therefore fully analogous to the two-level system as shown in Fig.~\ref{fig:rwa_ratio}.

\section{Application to many-body systems} \label{sec:results_many_body}

We now show that many-body systems can be cooled employing the same technique of randomized repeated interactions with the meter qubits. This generalization from the single-qubit case is possible despite several additional difficulties that arise for cooling many-body systems:
\begin{enumerate}
    \item \textit{Multiple transition frequencies:} Unlike a qubit, where there is a single transition frequency between the ground and excited states, many-body systems have multiple levels and transition frequencies that need to be addressed by the measurement qubits.
    \item \textit{Symmetry-protected subspaces:} For many-body systems, there may be symmetry-protected subspaces in the system's Hilbert space. When $H_\text{S}$ is unknown, the symmetries it supports are also unknown, making it difficult to break the symmetries in order to make protected subspaces accessible to cooling.
    \item \textit{Frustration:} Frustrated many-body systems present additional challenges for cooling to the ground state by \textit{(quasi-)local} system-meter couplings \cite{ticozzi_2012_Phil.Trans.R.Soc.A._StabilizingEntangledStates, wang_2025_StateEngineeringUnsteerable}. Frustration arises when the local interaction terms in $H_\text{S}$ cannot be minimized simultaneously, such that minimizing one forces another away from its minimum. Cooling protocols that attempt to lower the energy of the individual competing quasi-local terms may still reduce the total energy, but reaching the ground-state energy is challenging if not altogether impossible.
\end{enumerate}

\subsection{Cooling a Heisenberg chain} 
\label{subsec:Heisenberg}

In order to discuss a class of systems that feature all of the above, we consider a Heisenberg chain of size $N$ with the Hamiltonian
\begin{align}
    H_\text{S} &= \sum_{i=1}^{N-1} \sum_{a=x,y,z} J_{a} \sigma_{a,i} \otimes \sigma_{a,i+1},
   \label{eq:Anisotropic_Heisenberg_model}
\end{align}
where $J_a<0$ ($J_a>0$) corresponds to (anti-)~ferromagnetic coupling between sites. Conserved quantities are the total spin operators $\Sigma_{x,y,z} = \sum_i^N \sigma_{x,y,z,i}$ and $\Sigma^2 = \Sigma_x^2 + \Sigma_y^2 + \Sigma_z^2$.
In our cooling protocol, each system spin is coupled with a meter qubit: 
\begin{align}
    H_{\text{M}} &= \sum_i \frac{\omega_{i,\text{M}}}{2} \tau_{z,i}\,, \\
    H_\text{SM} &= \frac{\gamma}{2} \sum_i A_i (\theta, \phi) \tau_{x,i}\,. \label{eq:HSM}
\end{align}
In a single iteration of the protocol, all meter qubits have identically and independently distributed level splittings $\omega_{i,\text{M}}$, and each system spin is coupled to its corresponding meter qubit via a coupling operator $A_i$, which is also independently and identically distributed.
The steady-state energy of the system is calculated from the CPTP map $\bar{\mathcal{V}}$, averaged over the interactions $A_i(\theta,\phi)$ on the Bloch sphere, and meter splittings are sampled uniformly from the interval $\omega_\text{M} \in [0, 1.1 \mnorm{H_\text{S}}]$, requiring only knowledge of $\mnorm{H_\text{S}}$.

Figure~\ref{fig:heisenberg_3panels_a}-\ref{fig:heisenberg_3panels_c} displays the steady-state energy of the system for $N=3$. This is an isotropic antiferromagnetic Heisenberg chain with open boundary conditions, defined in Eq.~\eqref{eq:Anisotropic_Heisenberg_model}, with $J_a = (J,J,J)$, which serves as an example of a generic many-body frustrated model \cite{wang_2025_StateEngineeringUnsteerable}. The steady-state energy behaves qualitatively similarly to that in the single-qubit system, cf. Fig.~\ref{fig:tls_3panels}. In particular, it approaches the ground-state energy within an error of about $10^{-3} J$ for large $t_M$ and small $\gamma$ and
follows the estimate \eqref{eq:es_estimate} for weak coupling and long interaction times. This is in line with our interpretation of an effective RWA where resonant co-rotating interactions leading to cooling dominate over all other processes. 
Apparently, cooling by repeated randomized measurements can overcome the complications that emerge in many-body systems, which we identified above. We rationalize this as follows:
\begin{enumerate}
    \item 
Regarding the multitude of transition frequencies, sampling $\omega_\text{M}$ from the interval $[0, 1.1 \mnorm{H_\text{S}}]$ ensures that eventually a value is drawn, which is ``sufficiently resonant'' to any given transition in the system. Off-resonant values of $\omega_\text{M}$ are not detrimental, as the effective coupling between the system and meters is small in such cases.
Crucially, even if the ``wrong'' values of $\omega_\text{M}$ (potentially leading to heating) are chosen more frequently, for weak coupling between system and meters, the heating rate is slower than the cooling rate that occurs for ``good'' values of $\omega_\text{M}$. This is because sufficiently weak coupling and long interaction time realize the RWA (as illustrated in Fig.~\ref{fig:rwa_ratio} for the two-level system), and after many repetitions, one can effectively cool down the system. Here, it is important to keep in mind that the meters, prior to interaction, are prepared in their ground state. 
\item 
The issue of symmetries and symmetry-protected subspaces is resolved by randomizing the type of system-meter interactions. Here, resonant coupling to a meter qubit can be seen as a quasi-local perturbation that breaks the symmetry, and sampling the type of interaction randomly ensures that this will happen eventually, thus opening the protected subspace to cooling. 
Our procedure can be seen as a stochastic variant of quantum digital cooling where the system coupling operators $A_i$ are iterated over a prescribed set of operators~\cite{polla_2021_Phys.Rev.A_QuantumDigitalCooling}.
At the same time, for interactions that do not break the symmetry, cooling within the protected subspaces is still possible. That being said, breaking symmetries with randomly chosen interactions may be slow, and the presence of symmetries will likely reduce the cooling speed.
\item 
Cooling systems with frustrated Hamiltonians (``unsteerable systems'') requires a non-local action on the system. In our protocol, it is generated by the interplay of the system Hamiltonian and coupling to the meters when generating the time evolution. Even though the couplings within the Hamiltonian and between the system spins and meters are local (two-body in our example), the combined influence of local Hamiltonian evolution and repeated local resets leads, for sufficiently long interaction times $t_M$, to effectively non-local dynamics. In other words, the propagation of excitations governed by $H_\text{S}$ effectively dresses the local Kraus operators introduced by $H_\text{SM}$.
The overall protocol thus realizes a non-local action on the system, which may overcome frustration, similar to the circuit implementations of dissipative state preparation~\cite{ding_2024_Phys.Rev.Research_SingleancillaGroundState, fogarty_2016_Phys.Rev.A_OptomechanicalManybodyCooling}.
\end{enumerate}

The scaling of the steady-state energy with $t_M$ of the isotropic antiferromagnetic Heisenberg model obeying translation-invariance and open boundary conditions is shown for system sizes $N=3$ to $N=6$ in Fig.~\ref{fig:heisenberg_scaling}, revealing the energy to be proportional to $1/t_M$, as also expected from the estimate~\eqref{eq:es_estimate}. Already for $N=6$, the isotropic antiferromagnetic Heisenberg model exhibits a nontrivial spectral structure, with a high density of allowed transitions and pronounced degeneracies, see Appendix \ref{app:spectrum} for details, and thus captures key features of quantum many-body systems.
The steady-state energies in Fig.~\ref{fig:heisenberg_scaling} were obtained by averaging the CPTP map over the meter splitting and the system–meter interaction using Monte Carlo integration, taking the sampling size sufficiently large such that the residual statistical error is negligible compared to the reported energy values. The steady-state energy scaling is compared with the lower bounds on the energy that can be obtained by means of steering \cite{wang_2025_StateEngineeringUnsteerable}, indicated by dashed lines. Cooling by repeated randomized system-meter interactions clearly outperforms steering for sufficiently large $t_M$ and weak $\gamma$ in view of the scaling of the steady-state energy with $\gamma$ and $1/t_M$. This is discussed in more detail below. 

\subsection{Comparison with steering}\label{subsec:results_steering}

We compare cooling via repeated randomized system-meter interactions to measurement-induced steering, which prepares desired quantum states using specifically engineered quasi-local system-meter interactions~\cite{roy_2020_Phys.Rev.Research_MeasurementinducedSteeringQuantum}. 
Passive or ``blind'' steering, discarding measurement readouts, works perfectly for non-frustrated systems, where the target state is the ground state of a parent Hamiltonian (for which all its local terms individually annihilate its ground state), even in the absence of the system Hamiltonian. 
In the presence of the system Hamiltonian, the same steering protocol amounts to cooling the system to its ground-state energy. Provided the system is steerable, the ground state is reached with an error that decays exponentially in time.
When the meter reset rate is sent to infinity (corresponding to the continuous limit of steering), the dynamics of the steered system is described by a Lindblad master equation,
\begin{equation}
    \frac{d}{dt}\rho = -i [H_\text{S}, \rho] + \kappa \sum_i \left( L_i \rho L_i^\dagger - \frac{1}{2}\{L_i^\dagger L_i, \rho\} \right)\,.
    \label{eq:steering_master_equation}
\end{equation}
Here, the quasi-local jump operators $L_i$ are determined by the system-meter Hamiltonian~\cite{roy_2020_Phys.Rev.Research_MeasurementinducedSteeringQuantum,langbehn_2024_PRXQuantum_DiluteMeasurementInducedCooling}, and $\kappa$ is the steering rate given in terms of the coupling strength to the meters.

\begin{figure*}
    \centering
    \includegraphics[width=1\linewidth]{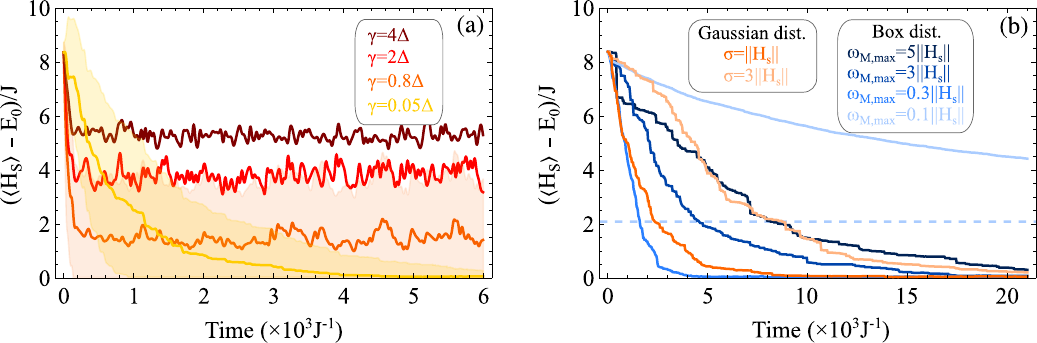}
    \vspace{-1\baselineskip}
    \caption{\textbf{Implementation of the cooling protocol:} The normalized energy expectation value of an $N=4$ anisotropic Heisenberg chain with $J_a= \{J, 1.2J, 0.7J\}$, cf. Eq. \eqref{eq:Anisotropic_Heisenberg_model}, as a function of the time. 
    (a) Dependence on the coupling strength $\gamma$: Weak coupling (with respect to the many-body gap $\Delta=2.134J$) guarantees cooling towards the ground state of the system (yellow line). The shaded regions depict the standard deviations around the mean values for the yellow and black curves; for the rest of the curves the standard deviations are omitted for clarity. The meter splittings are sampled from a box distribution with range from $0$ to $\omega_{\text{M},\max}=\left \Vert H_\text{S} \right \Vert$. 
    (b) Sampling of the meter splittings with $\gamma=0.04\Delta$ where box distributions (blue) cover $0$ to $\omega_{\text{M},\max}$ and Half-Gaussian distributions are centered at $0$ with variance $\sigma$ and sampling only $\omega_\text{M}\geq0$ (black) : Cooling is possible irrespective of the precise width of the box except for very narrow boxes (lightest blue, with the horizontal dashed line indicating the steady state energy). 
    For sufficiently broad box distributions, the width only affects the cooling rate. 
    Full agnosticity can be ensured by Gaussian distributions as their tails cover the entire range of possible system transitions.
    }
    \label{fig:implementation}
    \refstepcounter{subfigure}
    \label{fig:implementation_energy_gamma}
    \refstepcounter{subfigure}
    \label{fig:implementation_spectral_radius}
\end{figure*}

If the system under consideration is frustrated, the effectiveness of steering depends on the class of models to which it belongs \cite{wang_2025_StateEngineeringUnsteerable}. First, there is a family of non-frustration-free jittery steerable Hamiltonians, where steering may lead the system towards a multi-dimensional manifold. Once the system has reached this manifold, continuation of the protocol may induce  jumps among different states in this manifold. Second, there is a broad family of frustrated Hamiltonians (e.g., antiferromagnetic Heisenberg model, Dirac fermion and Majorana fermion SYK models, or the Fermi-Hubbard model) that can be cooled down to some degree using steering techniques, though not all the way to the ground state. In these cases, passive local steering is limited by the so-called ``glass floor'', which sets a lower bound on the achievable energy~\cite{wang_2025_StateEngineeringUnsteerable}. The term derives from an analogy to glassy systems featuring a rugged energy landscape, which are hard to cool. Steering typically cannot go beyond this energy threshold, because it effectively aims at minimizing energies associated with the quasi-local terms that make up the Hamiltonian $H_\text{S}$. But in frustrated systems, these terms do not have the global ground state as a common eigenstate with minimal energy. The value of the glass floor depends on the specific choice of the jump operators $L_i$. A lower bound on the glass floor can be obtained through a surrogate state construction \cite{wang_2025_StateEngineeringUnsteerable}, correspondingly referred to as  \emph{surrogate bound}. In Fig.~\ref{fig:heisenberg_scaling}, the surrogate bound for the Heisenberg antiferromagnet is shown by the dashed lines. The figure demonstrates that cooling can undercut the surrogate bound and, therefore, randomized cooling wins over any attempt at cooling by steering for sufficiently large $t_M$ and weak $\gamma$.

Finally, the system-meter interactions required to obtain the jump operators $L_i$ for steering are typically non-trivial to engineer. For systems with nearest-neighbor interactions, for instance, one needs to engineer three-body interactions between two system sites and the meter qubit~\cite{langbehn_2024_PRXQuantum_DiluteMeasurementInducedCooling}. This complication is absent in the proposed cooling procedure, where two-body interactions that do not require any fine-tuning are sufficient. Thus, our nature-inspired approach for cooling quantum matter provides a versatile framework that, in general, is more efficient than steering-based cooling, both from an abstract and a practical perspective. 

\subsection{Implementation of the cooling protocol} \label{subsec:implementation}

To demonstrate how our cooling protocol works in practice, we calculate the time evolution of the energy of an anisotropic Heisenberg chain, given by Eq. \eqref{eq:Anisotropic_Heisenberg_model}.
Sampling the interaction directions according to Eq.~\eqref{eq:HSM} and the meter splittings from either a box or Gaussian distribution, we calculate the time evolution $\rho_\text{S}\left(\vec{r}_n,\ldots,\vec{r}_1\right)$ following Eq.~\eqref{eq:single_trajectory}.

Figure~\ref{fig:implementation} confirms the conclusions drawn from the steady state energies in sections \ref{subsec:results_qubit} and \ref{subsec:Heisenberg}. In particular, it shows that cooling works for weak coupling (yellow line in Fig.~\ref{fig:implementation_energy_gamma}) with the system  reaching arbitrarily close to the ground state (whose energy we denote by $E_0$) after sufficiently many meter resets (overall 200 resets were used in the simulation, with $t_M=30J^{-1}$). The standard deviation of the system energy (shaded region) also gradually vanishes in the weak coupling regime, indicating that fewer and fewer excited states are populated.
In contrast, for intermediate or large $\gamma$, both energy (black and red lines) and standard deviation (not shown for strong coupling for clarity) remain finite, indicating that the system is in a mixed state with finite occupation of excited energy levels.

The role of the distribution from which the meter splittings are sampled is illustrated in Fig.~\ref{fig:implementation_spectral_radius}. It shows that a rough estimate of the spectral radius $\|H_S\|$ is sufficient for convergence towards the steady state when sampling from a box distribution $\omega_\text{M} \in \left[0, \omega_{\text{M},\max}\right]$ (cf. dark blue curves in Fig.~\ref{fig:implementation_spectral_radius}). Convergence breaks down, as one would expect, for too narrow widths --- when the box width is much smaller than $\left \Vert H_\text{S} \right \Vert$ (lightest blue curve in Fig.~\ref{fig:implementation_spectral_radius}), the distribution does not include splittings that match important system transitions that need to be addressed in order to cool down the system. This is reflected also in the steady-state energy of the very narrow box (dotted horizontal light-blue line in Fig.~\ref{fig:implementation_spectral_radius}). In contrast, for sufficiently large boxes, the width only influences the convergence speed.
While an order of magnitude estimate of $\left \Vert H_\text{S} \right \Vert$ is helpful to achieve fast convergence, box distributions with widths much larger than $\left \Vert H_\text{S} \right \Vert$ still guarantee convergence, at the expense of larger cooling times. 

A rough estimate of the spectral radius is typically not hard to obtain. For the example of Heisenberg chains, the bound $\left \Vert H_\text{S}\right\Vert \leq N \abs{J} \max_i \left\Vert H_i \right\Vert$, where $\left\Vert H_i \right\Vert$ are the spectral radii of the local Hamiltonians $H_i$, can serve this purpose. 
On the other hand, a distribution with a long tail is guaranteed to cover all relevant frequencies in the system, allowing for full system agnosticity without even assuming an estimate of the spectral radius. As an example, we consider a Gaussian distribution, sampling $\omega_\text{M}>0$ with probability $p(\omega_\text{M}) = \frac{2}{\sqrt{2\pi}\sigma}e^{-\frac{1}{2}\omega_\text{M}^2 / \sigma^2}$ (black curves in Fig.~\ref{fig:implementation_spectral_radius}).  Similar to the box distribution, the width of the Gaussian influences the rate of convergence but does not change the steady state. 
Figure~\ref{fig:implementation} thus confirms that cooling is realized for weak system-meter couplings and sampling of the meter splittings from either a boxed distribution with an order of magnitude estimate of the system's spectral radius or a tailed distribution agnostic of the system details.

\begin{figure}
    \centering
    \includegraphics[width=0.7\linewidth]{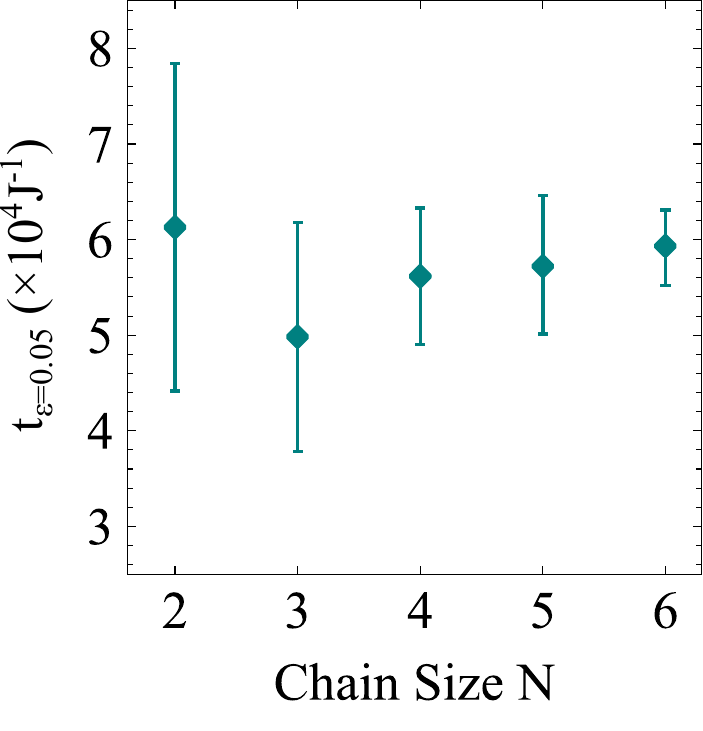}
    \vspace{-1\baselineskip}
    \caption{\textbf{Scaling of cooling time:} Time it takes to cool an anisotropic Heisenberg chain with $J_a= \{J, 1.2J, 0.7J\}$, cf. Eq. \eqref{eq:Anisotropic_Heisenberg_model}, down to an error $\varepsilon=0.05$ as a function of the system size. The error is quantified as $\varepsilon=(\langle H_S \rangle- E_0)/J$, where $E_0$ is the energy of the ground state. The values of the interaction time and the interaction strength are $t_M=80J^{-1}$ and $\gamma=0.03 J$, respectively. The meter splittings are sampled from a box distribution with range from $0$ to $\omega_{\text{M},\max}=10J$.}
    \label{fig:scaling_time}
\end{figure}

The decrease in energy variance observed in Fig.~\ref{fig:implementation_energy_gamma}~\footnote{More precisely, the standard deviation instead of the variance is plotted in order to match the units of energy.} can be leveraged for practical use: Under the paradigm of agnosticity, the spectrum of the system is not known and so measurements of the systems energy can not be used to infer success, failure, or convergence of the protocol. By contrast, an experimentalist with knowledge of $\var \left( \langle H_S \rangle \right)$ may use its vanishing as indicator of protocol success. To be explicit, although any (excited) energy eigenstate of energy $\epsilon$ shows $\var \left( \langle H_S \rangle \right)=0$, the protocol is designed such that by randomizing the interactions, eventually $H_\text{SM}$ will not commute with $H_\text{S}$ in the relevant eigenspace of energy $\epsilon$. At this point the system state and accordingly its energy and $\var \left( \langle H_S \rangle \right)$ will change. As explained in Sec.~\ref{subsec:RWA_qubit} and Appendix~\ref{app:ground_state_steady_state}, heating processes are suppressed such that the state after this interaction is lower in energy. Repeating the process, we approach the ground state as shown in Appendix~\ref{app:ground_state_steady_state}. So in consequence, the only state with $\var \left( \langle H_S \rangle \right)=0$ that is also steady is the ground state. In practice, one shall observe the variance and decrease, respectively, increase $\gamma$ and $t_\text{M}$ accordingly until convergence is reached.

In addition to the basic implementation of the protocol, a relevant question for applications is its scaling with system size.
Figure~\ref{fig:scaling_time} displays the time to reach the threshold $\varepsilon=0.05$ for the anisotropic Heisenberg model as a function of $N$, where $\varepsilon=(\langle H_S\rangle - E_0)/J$ quantifies the excess energy above the ground state energy (in units of $J$). Cooling times are averaged over 100 stochastic realizations, all starting from the initial state with all spins up. The error bars indicate the standard deviations and are found to be converged already for less than 70 realizations. Within the stochastic error bounds, no discernible increase of the cooling time with system size is observed in Fig.~\ref{fig:scaling_time}. This can be rationalized as the result of two competing size-dependent effects. On the one hand, increasing the system size tends to slow down the cooling dynamics since the ground state is a non-local state. On the other hand, since each site of the chain is coupled to a meter, increasing the system size simultaneously increases the total number of meter qubits employed in the protocol, which accelerates the removal of excitations from the chain. At the same time,  increasing $N$ may require an adjustment of $\gamma$ and $t_M$, to ensure a regime akin to the implementation of the RWA.

\begin{figure*}
    \centering
    \includegraphics[width=0.9\linewidth]{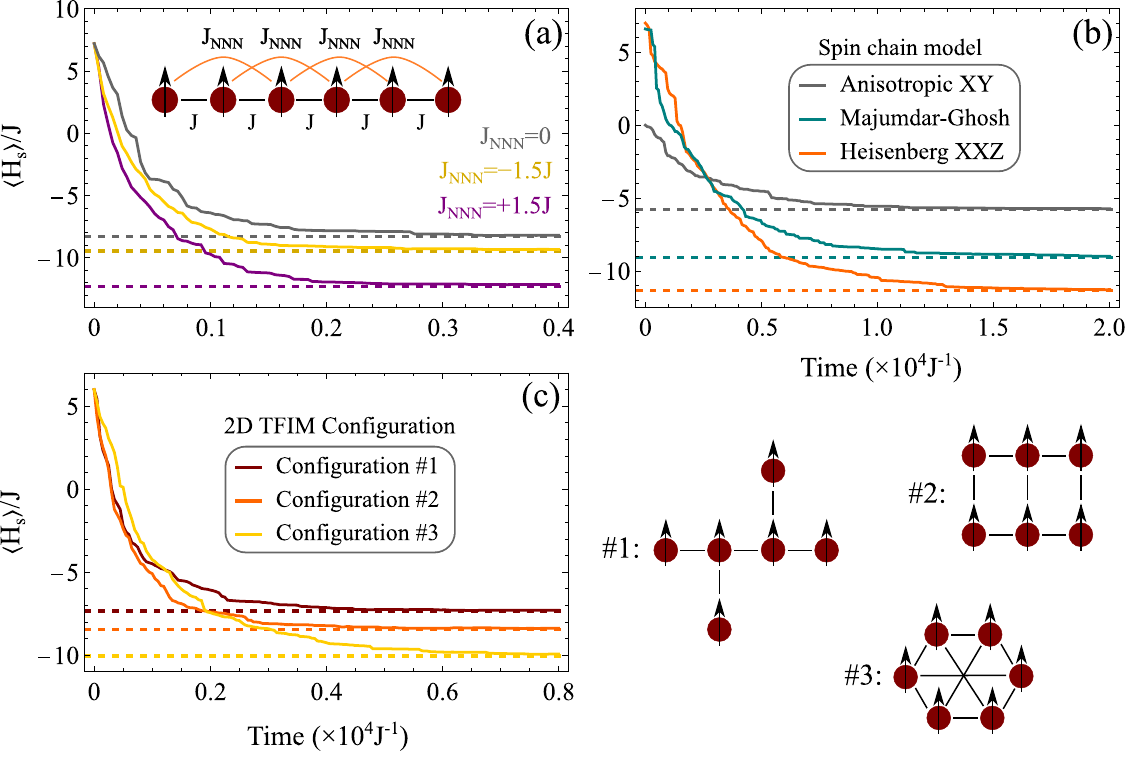}
    \vspace{-1\baselineskip}
    \caption{\textbf{Universality of the cooling protocol:} Energy expectation value as a function of time, approaching the ground state energy (marked by horizontal dashed lines) in all cases. (a) Transverse-field Ising model (TFIM) with next-nearest-neighbor couplings (also known as ANNNI model) for three values of the coupling strength $J_{\text{NNN}}$ with  a transverse field strength of $g=1.2J$,  system-meter interaction time $t_M=40J^{-1}$ and system-meter coupling $\gamma=0.08J$. (b) Anisotropic XY, Majumdar--Ghosh and Heisenberg XXZ model with anisotropy/relative-strength parameters  $\alpha=0.6$, $\beta = 0.4$, $\lambda= 1.4$, cf. Eq. \eqref{eq:Anisotropic_Heisenberg_model} with $J_a = \{ J, αJ, 0 \}$ and $J_a = \{ J, J, λJ\}$ for the XY and XXZ models, respectively, and Eq. \eqref{eq:MG_model} for the Majumdar--Ghosh model, $t_M=80J^{-1}$ and $\gamma=0.03J$. (c) TFIM in two-dimensional configurations with transverse field strength $g=J$,  $t_M=50J^{-1}$ and $\gamma=0.06J$. In all cases, $N=6$, the initial state of all models is all spins up, and the meter splittings were sampled uniformly from 0 to $\omega_{\text{M},\max}=10J$.}
    \label{fig:universality}
    \refstepcounter{subfigure}
    \label{fig:TFIM_NNN}
    \refstepcounter{subfigure}
    \label{fig:XY_MG_XXZ}
    \refstepcounter{subfigure}
    \label{fig:2D_TFIM}
\end{figure*}

To illustrate the universality of stochastic cooling, Fig.~\ref{fig:universality} shows its performance for several one- and two-dimensional spin models (with $N=6$). The transverse-field Ising model (TFIM) with next-nearest-neighbor (NNN) couplings, also known as the axial next-nearest-neighbor Ising (ANNNI) model, described by the Hamiltonian
\begin{equation} 
\begin{split} 
H_{\text{ANNNI}} = & g \sum_{i=1}^N σ_{z,i} - J \sum_{i=1}^{N-1} σ_{x,i} \otimes σ_{x,i+1} \\ & - J_{\text{NNN}} \sum_{i=1}^{N-2} σ_{x,i} \otimes σ_{x,i+2}. \label{TFIM_NNN} 
\end{split} 
\end{equation}
is featured in Fig.~\ref{fig:TFIM_NNN}. 
Convergence towards the ground state energy  (indicated by horizontal dashed lines) is observed for both positive and negative values of  $J_{\text{NNN}}$ as well as for $J_{\text{NNN}}=0$, in which case the model reduces to the standard TFIM.
We consider three additional spin chains, namely, the anisotropic XY Heisenberg model, the Majumdar--Ghosh model, and the XXZ Heisenberg model. The Hamiltonians of the XY and XXZ models are special cases of Eq. \eqref{eq:Anisotropic_Heisenberg_model} with $J_a = \{ J, αJ, 0 \}$ and $J_a = \{ J, J, λJ\}$, respectively, while the Hamiltonian of the Majumdar--Ghosh model is
\begin{equation} 
\begin{split}
   H_{\text{MG}} = & J \sum_{i=1} ^{N-1} \sum_{a=x,y,z} σ_{a,i} \otimes σ_{a,i+1} \\ &+\beta J \sum_{i=1} ^{N-2} \sum_{a=x,y,z} σ_{a,i} \otimes σ_{a,i+2}.  \label{eq:MG_model} 
\end{split}
\end{equation}
Here, $\alpha$ and $\lambda$ quantify the anisotropy between coupling strengths along different spin directions, while $\beta$ characterizes the relative strength of the next-nearest-neighbor to nearest-neighbor coupling. The cooling dynamics of the above Hamiltonians are studied in Fig. \ref{fig:XY_MG_XXZ}. The choice of these models was motivated by the fact that they capture representative features of quantum spin systems, such as frustration and strong quantum correlations. Taken together, all of the above-considered examples prove that our stochastic protocol is applicable across qualitatively different many-body Hamiltonians and enables cooling of all systems towards their ground states.
Finally, Fig. \ref{fig:2D_TFIM} illustrates the implementation of the stochastic protocol in two spatial dimensions, taking as example a TFIM in several two-dimensional configurations with   Hamiltonian
\begin{equation}
H_{\text{TFIM}}^{(2\text{D})} = g \sum_{i=1}^N σ_{z,i} - J \sum_{\langle i,j \rangle} \sigma_{x,i} \otimes \sigma_{x,j},
\end{equation}
where $\langle i,j \rangle$ denotes nearest-neighbor pairs on the 2D lattice. As expected, the geometry of the lattice does not impose any restrictions on the ability to cool the system towards its ground state.

\subsection{Limitations}\label{subsec:limitations}

The central result of this work is that universal cooling leads to a steady-state energy that can be arbitrarily reduced for large $t_M$ and small $\gamma$ according to Eq.~\eqref{eq:es_estimate}.
This estimate holds for weak coupling and long interaction times compared to the system's intrinsic energy scale. That is, a system can be cooled down as long as $\Delta/\gamma \gg 1$ and $\Delta t_M \gg 1$, where $\Delta$ is a typical transition energy in the system. This coincides with the regime of validity of the RWA, Eqs.~\eqref{eq:rwa1_weak_coupling}-\eqref{eq:rwa3_time}, supporting the idea that the resonant processes captured by the RWA provide the key mechanism for universal, i.e., system-agnostic, cooling.

A limitation to our protocol may arise from small energy gaps $\Delta$ which are not uncommon in many-body systems, especially close to the ground state. At first glance, this appears to be a potential obstacle for cooling: For $\Delta$ comparable to $\gamma$, the coupling to the meter may not be able to resolve the ground and first excited many-body levels. One would thus expect fluctuations between these two levels. This is indeed an obstacle for ground state preparation, where the figure of merit of the protocol is given in terms of a state-to-state distance measure. In contrast, cooling in terms of reducing the system energy or, equivalently, temperature is still possible: The steady state may in general be a mixed state of low excited states whenever $\Delta \sim \gamma$ and its energy expectation value is of the order of $\min\left(\Delta, \gamma\right)$. Thus the energy of the many-body system will be reduced until it is of that order.

Another limitation may arise from imperfectly prepared meters whereas we have assumed all meters to be accurately prepared in their ground state, i.e., a zero temperature bath. If, in practice, the meters are not perfectly initialized in their ground state but some thermal state, the system will thermalize to the temperature of the meter ensemble \cite{Englert2002}.
For a single-qubit system, we have investigated the implications of a meter prepared in a thermal state with finite temperature in Appendix~\ref{App:steering}. We have found that the occupation number of the meter gets transferred to the system's steady state. This confirms that the effective temperature found in the system's steady state is limited by the temperature of the meter, as one would expect.

\section{Discussion and outlook}\label{sec:conclusion}

We have shown that combining the well-established technique of repeatedly resetting auxiliary degrees of freedom~\cite{terhal_2000_Phys.Rev.A_ProblemEquilibrationComputation} with stochastic system-meter interactions results in a system-agnostic cooling protocol. 
Stochastic system-meter interactions imply that both heating and cooling processes can occur in the various steps of the protocol. We have found that weak and sufficiently long-time system-meter interactions ensure that ``good'' cooling interactions are much more efficient in extracting energy from the system than ``bad'' heating interactions that pump energy into the system. The reason is that the heating processes are driven by off-resonant interactions, and weak system-meter interactions make sure these are negligible. Such an effective implementation of the rotating-wave approximation realizes a non-trivial refrigerator that operates by disposing of the auxiliary degrees of freedom after they interact with the system. In other words, the rotating-wave approximation forms the basis for universal cooling. This is the key insight that our work provides. 

Our protocol is universal in the sense that lowering the system's energy is guaranteed when the strength of the system-meter coupling is reduced and the inverse interaction time becomes smaller than the minimum many-body energy gap. Thus, for many-body systems with non-degenerate ground states, preparation of the ground state with arbitrary accuracy is possible \emph{in the asymptotic limit.}

At first sight, this may seem counterintuitive. Naively, a zero-temperature reservoir---an engineered ``fridge'' with all auxiliaries prepared in their ground state---should cool the system to zero temperature irrespective of any conditions on the protocol. The actual dynamics of stochastic cooling is, however, more intricate for two reasons. (i) Part of the energy involved in each cooling step is associated with the system-meter interaction, which may lead to transient heating of the system even for zero-temperature auxiliaries. (ii) The meters have nowhere to further irreversibly transfer the energy received from the system, so that excitations return to the system if the interaction is switched off at an unfavorable time.
Therefore, sequential interaction with the local individual auxiliaries is not identical to the effect of genuinely dissipative modes, even though the auxiliaries are decoupled and reset after the interaction step. 
While the protocol emulates ``natural cooling'', implementing universal agnostic cooling via ``disappearing'' meters is conceptually much richer than immersing the system into the zero-temperature bath formed by auxiliaries.

The simplicity of our approach sets it apart from steering protocols \cite{roy_2020_Phys.Rev.Research_MeasurementinducedSteeringQuantum,herasymenko_2023_PRXQuantum_MeasurementDrivenNavigationManyBody,ding_2024_Phys.Rev.Research_SingleancillaGroundState}
which require one to engineer non-trivial system-meter interactions.
Moreover, while steering efficiently prepares certain predesignated target states, paradoxically, it may give rise to heating when applied to a system whose microscopic parameters are not fully known~\cite{uppalapati_2024_AntithermalizationHeatingCooling}. 
When comparing our stochastic universal cooling to approaches that implement dissipative state preparation via circuit decompositions~\cite{polla_2021_Phys.Rev.A_QuantumDigitalCooling,ding_2024_Phys.Rev.Research_SingleancillaGroundState,lin_2025_APLComput.Phys._DissipativePreparationManybody,zhan_2025_RapidQuantumGround}, these protocols are expected to be faster and to thus require less resources in the form of meter qubits, while relying on similar or somewhat higher levels of knowledge about the system. 
Generally, a higher level of system knowledge correlates with better scaling of the protocol. For example, the LogSweep algorithm for the transverse-field Ising model of Ref.~\cite{polla_2021_Phys.Rev.A_QuantumDigitalCooling} utilizes meter energy sweeps over a range that covers all possible system transitions to achieve cooling in polynomial times, but this assumes at least an estimate of the energy gap. On the other hand, the bang-bang-type protocols utilizing short, strong interactions to achieve rapid and depth-efficient cooling that have also been discussed~\cite{polla_2021_Phys.Rev.A_QuantumDigitalCooling}, are susceptible to control errors and suffer from convergence issues as the cooling process may plateau at some residual excitation levels.
Similarly, the logarithmic nature of the LogSweep algorithm~\cite{polla_2021_Phys.Rev.A_QuantumDigitalCooling} as well as implementations of dissipative state preparation using a spectral filter function~\cite{ding_2024_Phys.Rev.Research_SingleancillaGroundState} may be problematic when approaching sub-gap states. Moreover, the ingenuity of these sophisticated protocols requires some degree of control to be exerted, in contrast to our approach, where randomness of all parameters involving the meters is expected to make the procedure robust---indeed, any potential error in the implementation can be viewed as a source for randomness. 

Our study opens the door to a host of challenging questions. An important issue is the scalability of our method. This concerns both the cooling time and its dependence on the system-meter coupling strength and interaction time, as the system size is increased, as well as the required number of meter qubits. 
Evidently, one should consider the possibility of a dilute-cooling variant of our protocol, analogously to dilute cooling with steering~\cite{langbehn_2024_PRXQuantum_DiluteMeasurementInducedCooling}. 
In terms of improving the cooling time, it could be advantageous to exploit the fact that sometimes a low-temperature thermal state can be reached faster than a
high-temperature one~\cite{ramon-escandell_2025_Phys.Rev.Research_ThermalStatePreparation}. Moreover, active-decision protocols, which use the readout sequences to design the further protocol steps, may accelerate the preparation of target states~\cite{herasymenko_2023_PRXQuantum_MeasurementDrivenNavigationManyBody, volya_2024_IEEETrans.QuantumEng._StatePreparationQuantum, morales_2024_Phys.Rev.Research_EngineeringUnsteerableQuantum, morales_2025_Phys.Rev.Research_ScalableActiveSteering}. Similarly, our stochastic cooling protocol is likely to benefit from leveraging knowledge acquired from the measurement outcomes~\cite{marti_2025_Quantum_EfficientQuantumCooling}. In other words, instead of simply discarding the meters, one may read them out after their interaction with the system. Meter excitations will signal ``good'' interactions. One can then adapt the distribution from which the meter splittings and interactions are drawn to speed up the cooling, without changing the basic setup implementing the protocol.

It will also be interesting to understand the performance of our method as a function of the energy landscape being more or less ``glassy'' \cite{wang_2025_StateEngineeringUnsteerable}. Will the system get trapped in a metastable energy minimum, blocking its access to the true minimum, or is stochasticity sufficient to overcome such scenarios? In this respect, it is important to note the difference between cooling towards the ground state energy and attaining significant overlap with the ground state (high fidelity). This issue is particularly pertinent when the ground state is degenerate. Finally, one may consider combining our cooling protocol with quantum simulation or quantum annealing, where the system Hamiltonian is slowly varied in time. Ideally the dynamics is adiabatic but in case of gap closings, excitations are inevitably generated that translate into errors of the quantum simulation or state preparation. Our protocol can be leveraged to correct such errors. Unlike approaches based on counterdiabatic driving~\cite{takahashi_2017_Phys.Rev.A_ShortcutsAdiabaticityQuantum}, universal cooling does not assume knowledge of the time-dependent gap or state, and it is simpler to implement than engineered dephasing which requires non-local couplings~\cite{sveistrys_2025_Quantum_SpeedingQuantumAnnealing}. We therefore expect universal cooling to become a versatile and simple-to-use tool for the control of many-body quantum systems far from equilibrium, as relevant in quantum computing and quantum simulation.

\begin{acknowledgments}

We acknowledge helpful discussions with Yi-Xuan Wang, Kyrylo Snizhko, Yaroslav Herasymenko, Lin Lin, Alexander Mirlin, and Shovan Dutta.
Financial support from the Deutsche Forschungsgemeinschaft (DFG) through the Collaborative Research Centre TRR183 (project B02) and grants SH 81/8-1 and GO 1405/7-1, from the German Federal Ministry for Education and Research (BMBF), Project No. 13N16200 and 13N16201 NiQ, from by the National Science Foundation (NSF-2338819)–Binational Science Foundation (BSF-2023666), and from Berlin Quantum (BQ), an initiative endowed by the Innovation Promotion Fund of the city of Berlin, is gratefully acknowledged.

\end{acknowledgments}

\appendix

\section{Repeated measurements are formally equivalent to coupling to a real bath} \label{app:equivalence_to_continuous_bath}
A sequence of system-meter interactions is known to be equivalent with system interacting with a bath consisting of all meters simultaneously~\cite{cattaneo_2021_Phys.Rev.Lett._CollisionModelsCan,ciccarello_2022_PhysicsReports_QuantumCollisionModels}. Here we extend this to randomized meters and system-meter interactions, that is we show that, on average, repeated measurements with meter qubits that are initialized in their ground state but with random level splitting and random system-meter interactions mimic the coupling to a continuum of bath modes.
We use the result of Sec.~\ref{subsec:equivalence}: The average of the CPTP map of a system undergoing random measurements is given by the CPTP map of a single system-meter configuration, which is then averaged over all possible configurations \eqref{eq:app_cptp_average}. 
We consider the system-meter configuration to be specified by the meter splitting $\omega_\text{M}$ and the system-meter interaction $H_\text{SM}$.

In the first step, we derive an explicit expression for the map \eqref{eq:app_cptp_average} to second order in $\gamma$, averaging over system-meter configurations. Thereafter, we follow the same procedure to find an expression for the CPTP map obtained for a system coupled simultaneously to all meters that entered the average in step 1.
Let the system and meter(s) be governed by
\begin{align}
    H_\text{tot} &= H_\text{S} + H_\text{SM} + H_\text{M}, \\
    H_\text{M} &= \frac{\omega_\text{M}}{2}\, \tau_z, \\ 
    H_\text{SM} &= \gamma\, A \otimes \tau_x,
\end{align}
where $A$ is the operator that couples the system to the meter. We rewrite $A$ in terms of eigenoperators w.r.t. $H_\text{S}$ as
\begin{align}
    A &= \sum_{\omega>0}\left[A(\omega) +A^\dagger(\omega)\right], \label{eq:eigenoperator_1}\\
    A(\omega) &= \sum_{\epsilon'-\epsilon=\omega} \Pi(\epsilon) A\, \Pi(\epsilon'), \label{eq:eigenoperator_2}
\end{align}
where $\Pi(\epsilon)$ is the projector onto eigenstates of $H_\text{S}$ with eigenenergy $\epsilon$. This notation facilitates going to the interaction picture w.r.t. $H_\text{S}+H_\text{M}$, and we obtain the interaction-picture Hamiltonian:
\begin{align}
    H_\text{I}(t)\!=\!\gamma\!\sum_{\omega>0}\!\left[A(\omega)e^{-i\left(\omega-\omega_\text{M}\right)t}\!+\!A^{\dagger}\!(\omega)e^{i\left(\omega+\omega_\text{M}\right)t}\right]\tau_+\!+\!\text{H.c.}\!.
\end{align}

Now, consider the Dyson series for weak couplings. To be precise, weak coupling means $\gamma\ll J_\text{S} \equiv\max_{1\leq i\leq n}\abs{\lambda_i \left(H_\text{S}\right)}$, i.e., one needs to compare the coupling strength to the spectral radius of the system's Hamiltonian. To first order in $\gamma/J_S$, the Dyson series becomes
\begin{align}
    U(t)=\mathbb{I}-i\int_{0}^{t}H_\text{I}\left(t'\right)dt'.
\end{align}
Using Eq.~\eqref{eq:app_cptp_from_u}, we find the CPTP map:
\begin{align}
    \mathcal{V}\left(t , \omega_\text{M}\right) &= U_{00}(t,\omega_\text{M})\bullet U_{00}^\dagger(t,\omega_\text{M})\notag \\
    &+ U_{10}(t,\omega_\text{M})\bullet U^\dagger_{10}(t,\omega_\text{M}), \\
    U_{00}(t,\omega_\text{M})&=\mathbb{I}, \\
    U_{10}(t,\omega_\text{M})&=\!-i\gamma t\sum_{\omega>0}\!\Bigg[\!A\left(\omega\right)e^{-\frac{i}{2}\left(\omega-\omega_\text{M}\right)t}\sinc\!\left(\frac{\omega-\omega_\text{M}}{2}t\!\right) \nonumber\\
    &+A^{\dagger}\left(\omega\right)e^{\frac{i}{2}\left(\omega+\omega_\text{M}\right)t}\sinc\!\left(\frac{\omega+\omega_\text{M}}{2}t\!\right)\Bigg],
\end{align}
where $\sinc(x)=\sin(x)/x$ is the unnormalized sinc function.
If the meter splitting at each iteration is drawn uniformly from the interval $[0,\omega_{\text{M},\max}]$, then the configuration-averaged CPTP map becomes
\begin{align}
    \bar{\mathcal{V}}(t) &= \mathbb{I}+\frac{1}{\omega_{\text{M},\max}}\int_0^{\omega_{\text{M},\max}} U_{10}(t,\omega_\text{M})\bullet U_{10}(t,\omega_\text{M})d\omega_\text{M} .
    \label{eq:cptp_dyson}
\end{align}

In the second step, we consider a system coupled to all meter configurations in the above average at the same time. While the system Hamiltonian is identical to that in step 1, the meter's and interaction-picture coupling Hamiltonians now take the form 
\begin{align}
    \check{H}_\text{M}&= \int_0^{\omega_{\text{M},\max}} \tau_z \left( \omega_\text{M}\right)\,   d\omega_\text{M}, \\
    \check{H}_{I}(t)&=\gamma\,\frac{1}{\omega_{\text{M},\max}}\sum_{\omega>0}
    \int_{0}^{\omega_{\text{M},\max}} \!\Big[A\left(\omega\right)\tau_{+}\!\left(\omega_\text{M}\right)e^{-i\left(\omega-\omega_\text{M}\right)t} \notag
    \\& + A^{\dagger}\left(\omega\right)\tau_{+}\left(\omega_\text{M}\right)e^{i\left(\omega+\omega_\text{M}\right)t}\Big]d\omega_\text{M}  +\text{H.c.}
\end{align}
where $\tau\left(\omega_\text{M}\right)$ are the operators acting on the meter with splitting $\omega_\text{M}$ in the ``bath'' of meters.
Invoking again the Dyson series to first order in $\gamma$ and using Eq.~\eqref{eq:app_cptp_from_u}, we find the CPTP map to be
\begin{align}
    \check{\mathcal{V}}(t) &= U_{\vec{0}\vec{0}}(t)\bullet U_{\vec{0}\vec{0}}^{\dagger}(t)+\sum_{i=1}^{N_{m}}U_{\hat{e}_{i}\vec{0}}(t)\bullet U_{\hat{e}_{i}\vec{0}}^{\dagger}(t), \\
    U_{\vec{0}\vec{0}}(t)&=\mathbb{I}, 
\end{align}
where $N_m$ is the number of bath modes and the sum should be appropriately replaced by an integral $$\sum_{i=1}^{N_m}\longrightarrow\int_0^{\omega_{\text{M},\max}} d\omega_\text{M}.$$ 
This yields
\begin{align}
    \check{\mathcal{V}}(t) &= \mathbb{I} + \frac{1}{\omega_{\text{M},\max}}\int_0^{\omega_{\text{M},\max}}\!d\omega_\text{M} \, \check{U}_{10}(t,\omega_\text{M})\bullet \check{U}_{10}^{\dagger}(t,\omega_\text{M})
\end{align}
with $U_{10}(t,\omega_\text{M})=\check{U}_{10}(t,\omega_\text{M})$, i.e., we obtain the same expression as in step 1.
In conclusion, the CPTP map obtained by performing an average over system-meter configurations randomly drawn at every measurement interval is formally equivalent to a system coupled simultaneously to the continuum of configurations. The latter describes coupling to a continuum of bath modes.
\section{The steady state is the ground state} \label{app:ground_state_steady_state}

In this section, we will show that, within our cooling protocol, the steady state of the system generically becomes the ground state in the limit of long interaction times $t_M \rightarrow\infty$.
Before presenting the proof, let us  recall that, generically, the CPTP map on the system $\mathcal{V}$, obtained from the joint unitary evolution generated by $H_\text{S} + H_\text{M} + H_\text{SM}$ by tracing out the state of the meter, will be contracting and therefore have a steady state unless $H_\text{SM}$ is trivial. 
In order to see this, we remind ourselves that any contracting CPTP map has a steady state, or, conversely, a CPTP map has no steady state \textit{iff} it is unitary and maps pure states to pure states. Now, we are interested in interactions $H_\text{SM}$ that are non-trivial in the sense that $H_\text{SM}$ acts non-trivially on both the system and the meter. In the simplest case considered in this work, we write $H_\text{SM} = A\otimes B$. This implies that under the joint unitary 
$$U=e^{-i\left(H_\text{S} + H_\text{M} + H_\text{SM}\right) t_\text{M}},$$ 
there exist some pure product state $\ket{\psi_0}_\text{M}\otimes\ket{\phi_0}_\text{S}$ that will be entangled through the action of $U$. Therefore, after tracing out the meter, the system's state is 
$$\rho_\text{S} = \tr_\text{M} U \ketbra{\psi_0}_\text{M}\ketbra{\phi_0}_\text{S}U^\dagger,$$ 
which is a mixed state, indicating the entanglement generated by $U$. We conclude that the CPTP map on S is contracting because it is not unitary and, therefore, has at least one steady state.

Let us also note that we are considering the steady state of the CPTP map $\bar{\mathcal{V}}$. This means that it is the state obtained after an infinite but discrete number of applications of $\bar{\mathcal{V}}$. As such, the state after $n$ applications is $\rho(n)=\bar{\mathcal{V}}^n \rho(0)$ at the corresponding time of the $n$-th meter reset $t=n t_\text{M}$. Therefore, the steady state extracted from $\bar{\mathcal{V}}$ should be considered a \emph{stroboscopic} steady state, i.e., for $n\rightarrow\infty$ it is the state the system assumes at $t=nt_\text{M}$. Even though the system may well be converged for sufficiently large $n$, it can still undergo fluctuations between meter resets, and one should only consider the system right after the meter has been reset and not during a period of system-meter coupling. These fluctuations between meter resets can be observed in Fig.~\ref{fig:knowledge_hierarchy}.

Now, to prove that the system's ground state is such a steady state, we consider again the Dyson series \eqref{eq:cptp_dyson} and apply it to some state $\rho$. For the sake of readability, we omit the average over $\omega_\text{M}$ for now and consider only a single coupling operator $A$. The average over $A$ can be performed at the end of the calculation.
The map yields:
\begin{align}
\rho'&=  \rho\\
 & 
+\sum_{\omega,\omega'>0}a_{-}\left(\omega,\omega_\text{M}\right)a_{-}^{*}\left(\omega',\omega_\text{M}\right)A\left(\omega\right)\rho A^{\dagger}\left(\omega'\right)\nonumber\\
 &
+\sum_{\omega,\omega'>0}a_{+}\left(\omega,\omega_\text{M}\right)a_{+}^{*}\left(\omega',\omega_\text{M}\right)A^{\dagger}\left(\omega\right)\rho A\left(\omega'\right)\nonumber\\
 & 
+\sum_{\omega,\omega'>0}a_{-}\left(\omega,\omega_\text{M}\right)a_{+}^{*}\left(\omega',\omega_\text{M}\right)A\left(\omega\right)\rho A\left(\omega'\right)\nonumber\\
 & 
+\sum_{\omega,\omega'>0}a_{+}\left(\omega,\omega_\text{M}\right)a_{-}^{*}\left(\omega',\omega_\text{M}\right)A^{\dagger}\left(\omega\right)\rho A^{\dagger}\left(\omega'\right), \nonumber
\end{align}
where the coefficients are given by
\begin{align}
    a_{\pm}\left(\omega,\omega_\text{M}\right)=&\gamma \,
    \frac{e^{\pm\frac{i}{2}
    \left(\omega\pm\omega_\text{M}\right)t}}{\omega\pm\omega_\text{M}}\sin\left(\frac{\omega\pm\omega_\text{M}}{2}t\right).
\end{align}
In the limit of a large measurement duration, we can replace them with 
\begin{align}
    a_\pm(\omega,\omega_\text{M}) = \gamma \, \omega_\text{M}(\omega\pm\omega_\text{M}).
\end{align}
Taking into account that $\omega_\text{M}>0$ and that the sum above runs only over positive $\omega,\omega'$, find that the expression simplifies to
\begin{align}
    \rho'=	\rho+\frac{\gamma^2}{\omega_{\text{M},\max}} \int_0^{\omega_{\text{M},\max}}A\left(\omega_\text{M}\right)\rho A^{\dagger}\left(\omega_\text{M}\right) d\omega_\text{M},
\end{align}
where we have introduced the average over $\omega_\text{M}$. Note that the Dyson series \eqref{eq:cptp_dyson} only yields a unitary operator to the corresponding order in $\gamma$. To correct for this, we renormalize the state:
\begin{align}
    \rho'\mapsto \frac{1}{\tr \rho'} \rho'.
\end{align}

We are interested in the steady state of this map. A steady state has to be an eigenstate of the second part of the expression, such that we seek
\begin{align}
    \frac{\gamma^2}{\omega_{\text{M},\max}} \int_0^{\omega_{\text{M},\max}}A\left(\omega_\text{M}\right)\rho A^{\dagger}\left(\omega_\text{M}\right) d\omega_\text{M} = \lambda \,\rho.
\end{align}
We now make a crucial assumption: For every excited eigenstate $H_\text{S}\ket{\epsilon'} = \epsilon' \ket{\epsilon'}$ with $\epsilon'>0$, there exists some $\omega_\text{M}$ such that the coupling operator $A$ couples it to some eigenstate lower in energy, i.e. $\bra{\epsilon}A\ket{\epsilon'} \neq0$ for all $\epsilon'>0$ and some $\epsilon<\epsilon'$. Under this assumption, the operator $A(\omega_\text{M})$ effects a jump between eigenstates of $H_\text{S}$ with transition energy $\omega_\text{M}$ from the energetically higher eigenspace to the lower one. 

We show that once the system is in its ground-state manifold, it is trapped there. To that end, consider a state with well-defined energy $\epsilon$ where taking $\epsilon=0$ corresponds to the ground state:
\begin{align}
    \rho & =\sum_{i}p_{i}\ketbra{\epsilon_{i}}.
\end{align}
We now analyze the integrand $A\left(\omega_\text{M}\right)\rho A^{\dagger}\left(\omega_\text{M}\right)$ above in two steps: First, we show that the $\omega_\text{M}=0$ contribution leaves the system energy unchanged, then we consider transitions between eigenspaces with $\omega_\text{M}>0$. In both cases, we need to compute

\begin{align}
    A\left(\omega_\text{M}\right)\rho A^{\dagger}\left(\omega_\text{M}\right) & =\sum_{i}p_{i}A\left(\omega_\text{M}\right)\ketbra{\epsilon_{i}}A^{\dagger}\left(\omega_\text{M}\right).
\end{align}
We focus on 
\begin{align}
    &A\left(\omega_\text{M}\right)\ketbra{\epsilon_i} A^{\dagger}\left(\omega_\text{M}\right)\notag \\
    &\ =  \sum_{\mu'-\mu=\omega_\text{M}}\sum_{\nu'-\nu=\omega_\text{M}}\Pi\left(\mu\right)A\Pi\left(\mu'\right)\ketbra{\epsilon_i}\Pi\left(\nu'\right)A\Pi\left(\nu\right)\notag \\
    &\ =  \sum_{\epsilon-\mu=\omega_\text{M}}\sum_{\epsilon-\nu=\omega_\text{M}}\Pi\left(\mu\right)A\ketbra{\epsilon_i}A\Pi\left(\nu\right)\notag \\
    &\ =  \Pi\left(\epsilon-\omega_\text{M}\right)A\ketbra{\epsilon_i}A\Pi\left(\epsilon-\omega_\text{M}\right).
\end{align}
Thus, we find
\begin{align}
    A\left(\omega_\text{M}\right)\!\rho A^{\dagger}\!\left(\omega_\text{M}\right)\! =\! \sum_i p_i \Pi\left(\epsilon-\omega_\text{M}\right)\!A\ketbra{\epsilon_i}\!A\Pi\left(\epsilon-\omega_\text{M}\right),
\end{align}
and we observe that $A\left(\omega_\text{M}\right)\rho A^{\dagger}\left(\omega_\text{M}\right)$ is supported only in the $\epsilon-\omega_\text{M}$ eigenspace. It vanishes if $\epsilon-\omega_\text{M}$ is not an eigenenergy of the system and in particular also when $\epsilon-\omega_\text{M}<0$. For $\omega_\text{M}=0$, $A\left(0\right)\rho A^{\dagger}\left(0\right)$ remains in its original $\epsilon$ eigenspace, its energy expectation value does not change.

Now consider a ground state $\epsilon=0$ and denote it with $\rho_{0}=\sum_{i}p_{i,0}\ketbra{0_{i}}$. The $\omega_\text{M}=0$ contribution can only change the state within the $\epsilon=0$ eigenspace but does not change its energy. Any $\omega_\text{M}>0$ contribution vanishes because $\epsilon-\omega_\text{M}<0$. We conclude that then
\begin{align}
    \rho_{0}'= & \rho_{0}+\sum_{i}p_{i,0}\Pi\left(0\right)A\ketbra{0_{i}}A\Pi\left(0\right).
\end{align}
After normalization, the system energy is computed to 
\begin{align}
    E_{\text{S}}'= & \text{tr}\left(H_{\text{S}}\rho_{0}'\right)=
    \text{tr}\left(H_{\text{S}}\rho_{0}\right)=0
\end{align}
and hence, the ground space is steady.

\section{Cooling and heating of a single qubit for the steering-type system-meter coupling}
\label{App:steering}
We derive an explicit form of the the system energy after the $n$-th iteration of the protocol for a single qubit system coupled to the meter via Eq.~\eqref{eq:h_xx_interaction}. The matrix representation of the total Hamiltonian takes the form 
\begin{align}
H_{\text{tot}}&= \frac{1}{2}\begin{pmatrix}\omega_\text{M}+\omega_\text{S} & & &\gamma\\
 & \omega_\text{M}-\omega_\text{S} & \gamma\\
 & \gamma & -\omega_\text{M}+\omega_\text{S}\\
 \gamma&  &  & -\omega_\text{M}-\omega_\text{S}
\end{pmatrix},
\end{align} where we assume $\omega_\text{M}>0$. We find its eigenvalues to be 
\begin{align}
\epsilon_{1,2} &= \pm \mu\equiv\pm\sqrt{\gamma^2 + (\omega_\text{S}+\omega_\text{M})^2}, \\
\epsilon_{3,4} &= \pm \nu\equiv\pm\sqrt{\gamma^2 + (\omega_\text{S}-\omega_\text{M})^2}\,.
\end{align}
At any stage of the cooling process, after resetting the meter, the system and meter are in a product state. Ideally, the meter qubit should be prepared in its ground state. Here, we also consider imperfectly prepared meters by assuming that the meter is initialized in a thermal state with occupation number $n_\text{M}$
\begin{align}
    \rho_\text{M} &= \frac{e^{-({\beta_\text{M} \omega_\text{M}}/{2})\tau_z}}{\tr e^{-({\beta_\text{M} \omega_\text{M}}/{2})\tau_z}} = \diag \left(n_\text{M}, 1-n_\text{M} \right), \\
    n_\text{M} &= \frac{1}{1+e^{\beta_\text{M} \omega_\text{M}}}\,.
\end{align}
After resetting the meter to $\rho_\text{M}$, system and meter are in the product state
\begin{align}
\rho(n) = \rho_\text{M}\otimes \rho_\text{S} (n),
\end{align}
where $n$ counts the number of past iterations of the process. The system's energy at the end of the $(n+1)$-th iteration is given by
\begin{align}
    E_\text{S}(n+1)&= \tr \left[H_\text{S} \rho_\text{S}(n+1) \right],\\
    \rho(n+1) &= e^{-iH_\text{tot} t_M} \rho(n)\, e^{i H_\text{tot} t_M}.
\end{align}
Notably, we find that $E_\text{S}(n+1)$ does not depend on any coherences in $\rho_\text{S}(n)$ but just on the system energy at the end of the $n$-th iteration. This allows us to write:
\begin{align}
    E_\text{S}(n+1)& = E_\text{S}(n) +E_{\mu}(n) + E_{\nu}(n),
    \label{eq:eS_recursive}
    \end{align}
where    
    \begin{align}
    E_{\mu}(n)&=\phantom{-}\frac{\gamma^2}{4\mu^2} \left(\frac{\widetilde{\omega_\text{S}}}{2} - E_\text{S}(n)\right) \sin^2\frac{\mu t}{2}, \label{eq:eS_mu}\\
    E_{\nu}(n)&=-\frac{\gamma^2}{4\nu^2} \left(\frac{\widetilde{\omega_\text{S}}}{2}+ E_\text{S}(n)\right) \sin^2 \frac{\nu t}{2} \,,
    \label{eq:eS_nu} 
    \end{align}
and we have introduced    
    \begin{align}
    \widetilde{\omega_\text{S}}& = \omega_\text{S} \left(1-2n_\text{M}\right).
\end{align}
Now consider the RWA regime \eqref{eq:rwa1_weak_coupling}-\eqref{eq:rwa3_time}, which leads to
\begin{alignat}{2}
    \mu \gg \nu & \qquad\text{for } \omega_\text{S} >0, \\
    \mu \ll \nu & \qquad\text{for } \omega_\text{S} <0.
\end{alignat}
We associate the co- and counter-rotating processes with
\begin{alignat}{3}
    E_\text{co}=E_\nu, & \qquad E_\text{counter} = E_\mu& \qquad\text{for } \omega_\text{S} >0, \\
    E_\text{co}=E_\mu, & \qquad E_\text{counter} = E_\nu& \qquad\text{for } \omega_\text{S} <0,
\end{alignat}
which can be verified by repeating the above calculation \emph{after} performing the RWA. We observe that the co-rotating process leads to a slow oscillation in the system energy associated with energy flow towards the meter. The counter-rotating terms, on the other hand, lead to an increase in both system and meter energy due to negative energy associated with $H_\text{SM}$. If the counter-rotating terms dominate, one may find that the system is heated instead of cooled. This is illustrated in Fig.~\ref{fig:rwa_ratio}.

One can solve the recursion \eqref{eq:eS_recursive} to find 
\begin{widetext}
\begin{align}
    E_{\text{S}}(n) =& \frac{\mu^{2}\left(E_{\text{S}}\left(0\right)+\frac{\widetilde{\omega_\text{S}}}{2}\right)\sin^{2}\frac{\nu t_{\text{M}}}{2}+\nu^{2}\left(E_{\text{S}}\left(0\right)-\frac{\widetilde{\omega_\text{S}}}{2}\right)\sin^{2}\frac{\mu t_{\text{M}}}{2}}{\mu^{2}\sin^{2}\frac{\nu t_{\text{M}}}{2}+\nu^{2}\sin^{2}\frac{\mu t_{\text{M}}}{2}}\left(1-\frac{\gamma^{2}}{\mu^{2}}\sin^{2}\frac{\mu t_{\text{M}}}{2}-\frac{\gamma^{2}}{\nu^{2}}\sin^{2}\frac{\nu t_{\text{M}}}{2}\right)^{n} \nonumber\\
    &-\frac{\widetilde{\omega_\text{S}}}{2}\frac{\mu^{2}\sin^{2}\frac{\nu t_{\text{M}}}{2}-\nu^{2}\sin^{2}\frac{\mu t_{\text{M}}}{2}}{\mu^{2}\sin^{2}\frac{\nu t_{\text{M}}}{2}+\nu^{2}\sin^{2}\frac{\mu t_{\text{M}}}{2}}. \label{eq:eS_recursive_sol}
\end{align}
\end{widetext}
Now, we focus on the limit $n\rightarrow \infty$. To that end, inspect the term $\left(\ldots \right)^n$. We find that with $\gamma^2 \leq \mu ^2, \nu^2$, its content lies in $[-1,1]$. Therefore, except for pathological choice of parameters of measure $0$, this term vanishes as $n\rightarrow\infty$. The asymptotic system energy becomes
\begin{equation}
    E_{\text{S}} \left(n\rightarrow\infty\right) = -\frac{\widetilde{\omega_\text{S}}}{2}\frac{\mu^{2}\sin^{2}\frac{\nu t_{\text{M}}}{2}-\nu^{2}\sin^{2}\frac{\mu t_{\text{M}}}{2}}{\mu^{2}\sin^{2}\frac{\nu t_{\text{M}}}{2}+\nu^{2}\sin^{2}\frac{\mu t_{\text{M}}}{2}}. \label{eq:eS_steady_state}
\end{equation}
This is the energy of the system in its steady state, and we show it in Fig.~\ref{fig:rwa_ratio}.

The solutions \eqref{eq:eS_recursive_sol} and \eqref{eq:eS_steady_state} are exact. Now, we focus again on the RWA parameter regime \eqref{eq:rwa1_weak_coupling}-\eqref{eq:rwa3_time}. To first order in the ratio $r = \nu / \mu$ for $\omega_\text{S}>0$ resp. $1/r$ for $\omega_\text{S} < 0$, the asymptotic system energy becomes
\begin{equation}
E_{\text{S}}\left(n\rightarrow\infty\right)=\frac{\widetilde{\omega_\text{S}}}{2}\begin{cases}
1-\dfrac{\mu^{2}t_{\text{M}}^{2}}{\mu^{2}t_{\text{M}}^{2}+2\sin^{2}\frac{\mu t_{\text{M}}}{2}}, & \omega_\text{S}>0,\\
\dfrac{\left(\frac{\nu t_{\text{M}}}{2}\right)^{2}-\sin^{2}\frac{\nu t_{\text{M}}}{2}}{\left(\frac{\nu t_{\text{M}}}{2}\right)^{2}+\sin^{2}\frac{\nu t_{\text{M}}}{2}}, & \omega_\text{S}<0,
\end{cases}
\end{equation}
If we choose the interval length $t_\text{M} \gg 1/\mu$ and $t_\text{M} \gg 1/\nu$ for $\omega_\text{S} >0$ and $\omega_\text{S} < 0$, respectively, the system energy becomes
\begin{equation}
    E_{\text{S}}\left(n\rightarrow\infty \right) = -\frac{\abs{\omega_\text{S}}}{2} \left(1-2n_\text{M}\right),
\end{equation}
which is the ground-state energy, irrespective of the sign of $\omega_\text{S}$ for perfectly prepared meters with $n_\text{M} = 0$. If the meter is prepared in a thermal state with finite inverse temperature $\beta_\text{M}$, we expect the meter's occupation number to be copied over to the system. That is, the system's steady state is $\rho_\infty = \rho_\text{M}$. The effective inverse temperature of the system associated with this state is
\begin{align}
    \beta_\text{S} = \beta_\text{M} \frac{\omega_\text{M}}{\omega_\text{S}}. \label{eq:imperfect_meter_temperature}
\end{align}
This indicates that for fixed $\omega_\text{M}$ close to resonance and fixed interaction, it is possible to drive the system to a state that exhibits a lower temperature than that of the bath of meters it is coupled to. It should be noted that Eq.~\eqref{eq:imperfect_meter_temperature} is only valid close to resonance such that the effect is small. Cooling beyond the limits of simple thermalization has been studied previously in \cite{deoliveirajunior_2025_Phys.Rev.Lett._HeatWitnessQuantum}.

\section{RWA: Perturbation theory, validity, and revival of counter-rotating terms} \label{app:rwa_validity}
Here, we address the RWA in more detail and demonstrate why the validity of the RWA is determined by Eqs.~\eqref{eq:rwa1_weak_coupling}-\eqref{eq:rwa3_time} of the main text.
We argue that it can be understood as an interference phenomenon where the magnitude of fast-oscillating (non-secular or counter-rotating) terms collapses when compared to the slow oscillating (secular or co-rotating) ones. Like with other interference phenomena, this implies that the counter-rotating terms are ``revived'' at some point in time (cf. collapse and revival of the excited-state population in the Jaynes-Cummings model when the cavity is in a coherent state). The purpose of this subsection is also to take the RWA beyond the folklore of saying ``fast oscillations average out''. In doing so, we focus on the situation at hand: two qubits coupled with an $\sigma_x \tau_x$ interaction via the Hamiltonian \eqref{eq:h_xx_interaction}.

The RWA is a perturbative approximation: The interaction strength $\gamma$ is assumed to be weak w.r.t. the bare system Hamiltonian. Its spectral radius is given by $(\omega_\text{S}+\omega_\text{M})/2$. Therefore, the first condition that needs to be satisfied is $\gamma \ll  (\omega_\text{S} + \omega_\text{M})/2$. In this case, we can treat the qubit-qubit coupling as a perturbation. Before doing so, we transform into the interaction picture w.r.t. the bare qubit Hamiltonians. We obtain
\begin{align}
    H_{\text{rot}}(t) &= \frac{\gamma}{2} \left[\sigma_+ \tau_+ e^{i\left(\omega_\text{S} + \omega_\text{M}\right)t/2} + \sigma_-\tau_+ e^{-i\left(\omega_\text{S}-\omega_\text{M}\right)t/2} \right] \notag \\
    &+\text{H.c.}.
\end{align}
For weak coupling, the time-evolution operator can be expressed to first order in $\gamma/\omega_\text{S}$ as
\begin{align}
    U_{I}\left(0,t\right) &= \mathcal{T} \exp{-i\int_{0}^{t} H_{\text{rot}}(t') dt'}\\
    &=\mathbb{I} -i \int_0^t H_{\text{rot}}(t')dt',
\end{align}
which is evaluated to 
\begin{widetext}
\begin{equation}
    U_{I}\left(0,t\right)=\mathbb{I}-i\frac{\gamma}{2}\,\left[\sigma_{+}\tau_{+}\frac{2e^{\frac{i}{2}\left(\omega_\text{S}+\omega_\text{M}\right)t}}{\omega_\text{S}+\omega_\text{M}}\sin\left( \frac{\omega_\text{S}+\omega_\text{M}}{2}t\right)+\sigma_{-}\tau_{+}\frac{2e^{-\frac{i}{2}\left(\omega_\text{S}-\omega_\text{M}\right)t}}{\omega_\text{S}-\omega_\text{M}}\sin\left(\frac{\omega_\text{S}-\omega_\text{M}}{2}t\right)+\text{H.c.}\right].
\end{equation}    
\end{widetext}
Let us assume for simplicity that $\omega_\text{S},\omega_\text{M}>0$ and that $\abs{\omega_\text{S}-\omega_\text{M}} \ll \omega_\text{S} + \omega_\text{M}$. In this case, the usual RWA procedure is to simply neglect terms oscillating with $\omega_\text{S}+\omega_\text{M}$ and retain those with $\omega_\text{S}-\omega_\text{M}$. We now want to take a closer look: Consider the ratio of the counter-rotating over co-rotating amplitudes
\begin{equation}
    r=\abs{\frac{\omega_\text{S} - \omega_\text{M}}{\omega_\text{S}+\omega_\text{M}}\frac{\sinc\left(\frac{\omega_\text{S}+\omega_\text{M}}{2}t\right)}{\sinc\left(\frac{\omega_\text{S}-\omega_\text{M}}{2}t\right)}} \,.
    \label{eq:rwa_ratio}
\end{equation}
If $r \ll 1$, then the RWA is justified. In the regime where $\left(\omega_\text{S} + \omega_\text{M}\right)/\abs{\omega_\text{S} - \omega_\text{M}} \equiv q\ll 1$, we can expand $r$ to leading order in $q$ to find
\begin{align}
    r =&\abs{\frac{\omega_\text{S} - \omega_\text{M}}{\omega_\text{S}+\omega_\text{M}} \sinc \left(\frac{\omega_\text{S}+\omega_\text{M}}{2}t\right)}+\mathcal{O}(q^3)\,.
\end{align}

As the magnitude of the function $\sinc(x)$ decays as $1/x$ for large $x$, we conclude that the RWA is then valid for times $t \gg 2/(\omega_\text{S}+\omega_\text{M})$. There is, however, a caveat: Although $\omega_\text{S}-\omega_\text{M}$ may be negligible when compared to $\omega_\text{S}+\omega_\text{M}$, as long as it is finite, it can play a role at large $t$. Indeed the denominator in Eq.~\eqref{eq:rwa_ratio} has roots at $t={4\pi n}/({\omega_\text{S}-\omega_\text{M}})$ with $n\in \mathbb{N}$. 
Let's take a step back and consider how the ratio of co- and counter-rotating terms evolves: At $t=0$ we have $\sinc\left(\frac{\omega_\text{S}+\omega_\text{M}}{2}t\right)=1$. Then in the interval $2/(\omega_\text{S} + \omega_\text{M})\ll t\ll 4\pi / (\omega_\text{S}-\omega_\text{M})$ the co-rotating terms dominate. We interpret this as a ``collapse'' of counter-rotating contributions. As soon as we hit $t=4\pi n / (\omega_\text{S}-\omega_\text{M})$, the counter-rotating terms are ``revived'' briefly before collapsing again. This process is depicted in Fig.~\ref{fig:rwa_ratio}.
Note that although the counter-rotating terms show a revival at certain times, their magnitude relative to the co-rotating terms is suppressed by $(\omega_\text{S}-\omega_\text{M})/(\omega_\text{S}+\omega_\text{M})$.

\section{Transition structure of the isotropic antiferromagnetic and the anisotropic XY Heisenberg spin chain}\label{app:spectrum}

\begin{widetext}

\begin{figure*}[h!]
    \centering
    \includegraphics[width=0.99\linewidth]{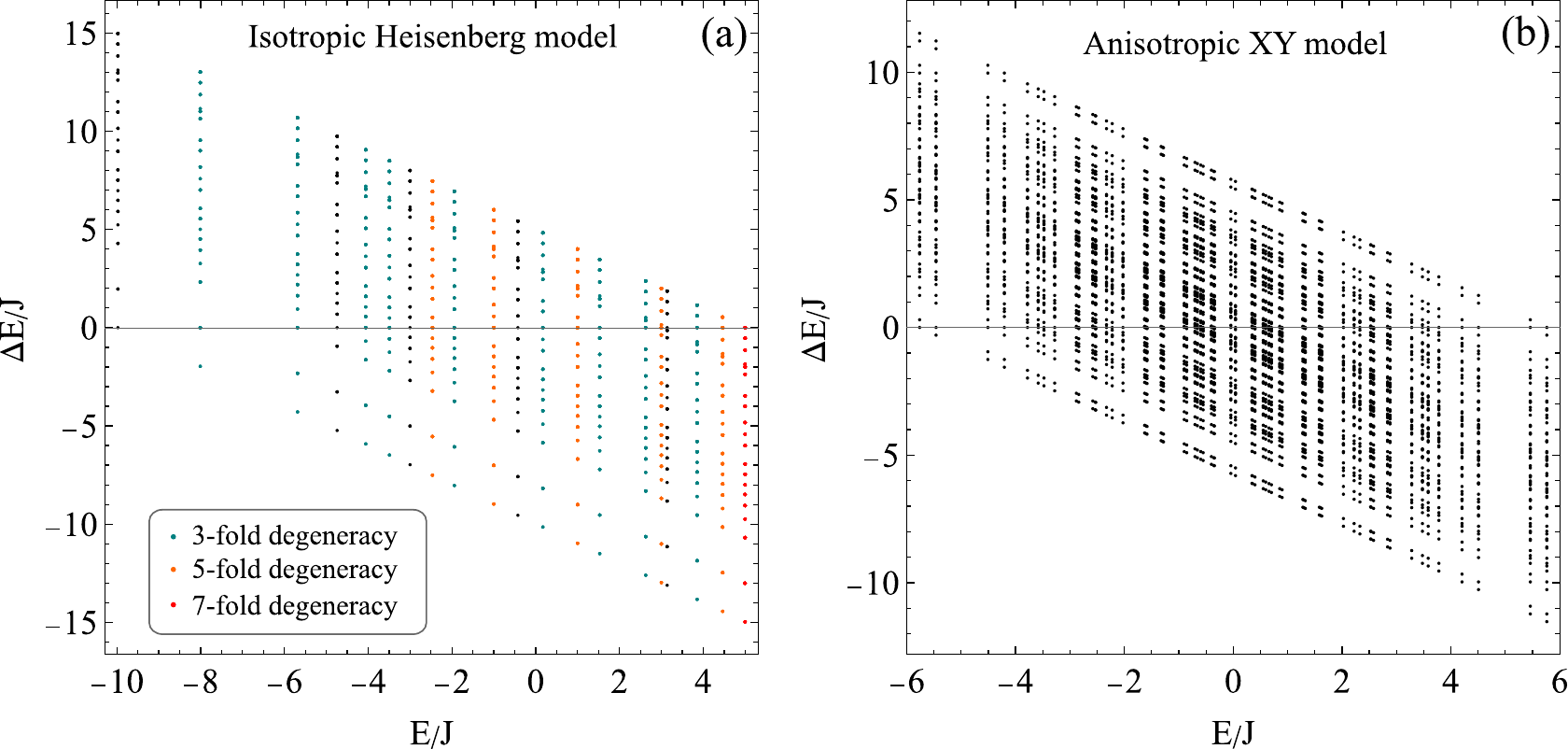}
    \vspace{-1\baselineskip}
    \caption{\textbf{Spectral properties of two spin-chain examples (with $N=6$):} Transition energies $\Delta E$ versus energy $E$ for (a) the isotropic antiferromagnetic Heisenberg chain (with color encoding the degeneracy of the levels) and (b) the anisotropic XY chain with anisotropy parameter $\alpha=0.6$, see  Eq.~\eqref{eq:Anisotropic_Heisenberg_model} with $J_a=\{J,\alpha J,0 \}$, and no degeneracies. 
    }
    \label{fig:Isotropic&XY_spectrum}
\end{figure*}
\end{widetext}

To illustrate the spectral complexity of the examples used with the stochastic cooling protocol in  \ref{subsec:implementation}, Fig. \ref{fig:Isotropic&XY_spectrum} displays the transition structure of an isotropic antiferromagnetic Heisenberg model, Eq.~\eqref{eq:Anisotropic_Heisenberg_model} with $J_a=\{J,J,J \}$, and an anisotropic XY chain, Eq.~\eqref{eq:Anisotropic_Heisenberg_model} with $J_a=\{J,\alpha J,0 \}$ (with $N=6$ for both), through the distribution of transition energies $\Delta E$ over the eigenenergies $E$ of the corresponding states. Already at this system size, both models exhibit a complex pattern of allowed transitions within a comparatively narrow energy range. In the case of the isotropic Heisenberg model, the complexity is largely associated with the presence of highly degenerate levels, including three-, five-, and seven-fold degeneracies, which reflect the symmetry of the system. In the case of the anisotropic XY model, the absence of degeneracies leads to a larger number of distinct eigenvalues and therefore a richer transition structure. These features highlight that the examples used in the main text capture essential aspects of a genuine many-body system already for very moderate system sizes such as $N=6$. Our examples thus provide appropriate settings for testing the stochastic cooling protocol that reflect the spectral complexity characteristic of many-body systems.

\clearpage

\bibliography{refs}

\end{document}